\documentclass[twocolumn,pre,floatfix,superscriptaddress,pre,showpacs]{revtex4-1}

\usepackage{amsmath}
\usepackage{amsfonts}
\usepackage{amssymb}
\usepackage{graphicx}
\usepackage{color}

\usepackage{graphicx,color}
\usepackage{amsmath,amssymb,amsfonts}
\usepackage{bm}
\usepackage{bibentry, natbib}
\usepackage[T1]{fontenc}
\usepackage[english]{babel}
\usepackage[latin1]{inputenc}
\usepackage{ifthen}
\usepackage{footnote}
\usepackage{subfigure}
\usepackage{setspace}
\usepackage{verbatim}
\usepackage{textcomp}

\definecolor{grey}{rgb}{0.5,0.5,0.5}
\newcommand{\be}{\begin{equation}}
\newcommand{\ee}{\end{equation}}
\newcommand{\rmd}{{\rm d}}
\newcommand{\rme}{{\rm e}}
 
\newcommand{\avg}[1]{\overline{#1}}              
\newcommand{\reff}[1]{(\ref{#1})} 

\newcommand{\change}[1]{\textcolor{black}{#1}}

\begin{document}

\title{Fixation properties of subdivided populations with balancing selection}
\author{Pierangelo Lombardo}
\affiliation{SISSA -- International School for Advanced Studies and INFN, via Bonomea 265, 34136 Trieste, Italy}
\author{Andrea Gambassi} 
\affiliation{SISSA -- International School for Advanced Studies and INFN, via Bonomea 265, 34136 Trieste, Italy}
\author{Luca Dall'Asta} 
\affiliation{Department of Applied Science and Technology -- DISAT, Politecnico di Torino, Corso Duca degli Abruzzi 24, 10129 Torino, Italy}
\affiliation{Collegio Carlo Alberto, Via Real Collegio 30, 10024 Moncalieri, Italy}

\date{\today}

\begin{abstract}
In subdivided populations, migration acts together with selection and genetic drift and determines their evolution. Building up on a recently proposed method, which hinges on the emergence of a time scale separation between local and global dynamics, we study the fixation properties of subdivided populations in the presence of balancing selection. The approximation implied by the method is accurate when the effective selection strength is small and the number of subpopulations is large. 
In particular, it predicts a phase transition between species coexistence and biodiversity loss in the infinite-size limit and, in finite populations, a nonmonotonic dependence of the mean fixation time on the migration rate.
In order to investigate the fixation properties of the subdivided population for stronger selection, we introduce an effective coarser description of the dynamics in terms of a voter model with intermediate states, which highlights the basic mechanisms driving the evolutionary process.
\end{abstract}

\pacs{05.40.-a, 87.23.Kg, 87.23.Cc, 05.70.Fh}

\maketitle
 
\section{Introduction}

In a natural population without any structure or subdivision
--- usually referred to as a \emph{well-mixed} population --- the temporal evolution results from the competition between the deterministic evolutionary forces (\emph{selection}) and the stochastic effects generated by the death and reproduction of individuals (\emph{genetic drift}). Because of stochasticity,  every finite population in the absence of mutations eventually reaches an absorbing state (\emph{fixation}) in which all individuals have a unique trait (e.g., species/language/opinion) and biodiversity, defined as the coexistence 
of various traits, is lost. When a population presents an internal structure, e.g., it is subdivided in subpopulations,  the individuals can move between different subpopulations and migration acts together with selection and genetic drift, influencing relevant long-time properties of the dynamics, such as the \emph{mean fixation time} (MFT). It is widely observed that habitat fragmentation and population subdivision play a major role in the process of ecological change and biodiversity loss.  Understanding and predicting the effects of migration on the collective behavior of a subdivided population is therefore of primary importance in order to preserve ecosystems and species abundance. 

Depending on the specific landscape into which the natural population is embedded, its spatial structure can be conveniently modeled by means of either one-, two-, three-dimensional regular lattices or, more generally, by a network with certain connections. 
If the degree of connectivity of each node of this network is sufficiently large and the connected subpopulations have constant and equal sizes, the effects of subdivision typically amount to a rescaling of the relevant parameters of the population, such as the effective population size $N_{\rm e}$ and the effective strength $s_{\rm e}$ of selection \cite{maruyama,slatkin}.
However, both $N_{\rm e}$ and $s_{\rm e}$ are functions of the rate with which individuals migrate between subpopulations, therefore both the fixation probability and the MFT  depend on it. 
\change{In the absence of selection or when it is constant,}
the MFT monotonically decreases upon increasing the migration rate \cite{whitlock,cherry-wakeley,blythe}.
When evolutionary forces which favor biodiversity are present, instead, it was recently shown that the MFT can display a nonmonotonic dependence on the migration rate \cite{prl}. 
Even in the absence of mutation, this kind of evolutionary forces are common to natural populations. In particular the {\em balancing selection} \cite{hamilton,balancing} associated with them is an umbrella concept, encompassing mechanisms such as over-dominance or heterozygote advantage, which act for the maintenance of biodiversity in several contexts, most 
notably mammalian \cite{hcc} and plants \cite{sil}. 
For example, it has been proposed that some genetic diseases in humans, such as sickle-cell anemia \cite{anemia}, cystic fibrosis \cite{fibrosis}, and thalassemia \cite{thalassemia} actually persist as a consequence of balancing selection.
Analogous mechanisms are responsible  for  the emergence of bilingualism in language competition \cite{abrams} or for cooperative behaviors in ecology and coevolutionary dynamics \cite{egt,traulsen}, such as those recently observed in microbial communities \cite{xavier}.

The present work builds on the approach introduced in Ref.~\cite{prl} and extends the investigation reported therein in several respects.
We consider a group of equally sized subpopulations (i.e., a \emph{metapopulation}) which balancing selection acts on, while migration takes place between any pair of subpopulations, such as to form a fully connected graph with each subpopulation occupying one of its vertices.  For concreteness we adopt below the terminology and  notation specific of population genetics. 
In Sec.~\ref{sec:metapopulation} we review the details of the model and the approximation proposed in Ref.~\cite{prl}, which hinges on the emergence of a 
separation between the time scale of the local dynamics occurring at each vertex of the network and that of the global dynamics at the level of the whole network. The resulting approximation turns out to be accurate when the effective selection strength is sufficiently small and the number $N$ of subpopulations is sufficiently large in the sense specified further below. 
In Sec.~\ref{sec:PT} we show that in this case a phase transition takes place between species coexistence and biodiversity loss.
In order to be able to investigate the fixation properties of the system for larger values of the selection strength --- which are not immediately accessible with the previous approach --- we propose in Sec.~\ref{sec:voter} an 
effective description in terms of a voter model, which is generically accurate for small values of the migration rate.
%
\change{This coarser description applies to a generic metapopulation model, independently of the specific form of the natural selection; in addition, in the presence of balancing selection, the range of values of migration rate for which the effective voter model provides accurate predictions can be extended 
by introducing an additional intermediate state in the voter model, as we discuss in Sec.~\ref{sec:voter_int}.} 
In contrast to the standard voter model, the one with the additional state is actually able to reproduce the distinctive nonmonotonic dependence of the MFT on the migration rate found in Ref.~\cite{prl}.
Moreover, it provides a semi-quantitative explanation of a nonmonotonic behavior observed in the MFT as a function  of the selection coefficient $s$, which appears for small migration rate $m$ in addition to the one discussed in Ref.~\cite{prl}.
A summary of our findings and the conclusions are then presented in Sec.~\ref{sec:conclusions}.

\change{\section{Metapopulation model}}
\label{sec:metapopulation}

\change{\subsection{From microscopic to mesoscopic dynamics}}
\label{sec:micro-meso}

The evolution of finite well-mixed populations is conveniently described at the microscopic level by the Wright-Fisher model \cite{sFisher,sWright}, 
which consists of a (haploid) population of $\Omega$ individuals, each one carrying one of two possible alleles $A$ or $B$. 
At each time step of the dynamics the original population is substituted by a new generation obtained by a binomial random sampling determined by the features of the previous one: the allele of each new individual is randomly drawn with a probability which depends on the frequency of occurrence of $A$ (or, equivalently $B$) \change{in the parent 
generation.} 
The time interval $\tau_g$ between two consecutive steps of this dynamics represents the duration of a generation. In a neutral model, i.e., in the absence of selection, each new individual carries allele $A$ (respectively $B$) with probability  $x=\Omega_A/\Omega$ (respectively $1-x$), where $\Omega_A$ is the number of individuals carrying allele $A$
\change{in the preceding generation.}
%
In order to mimic the effects of natural selection, one introduces different allele fitnesses  $w_A=1+\tilde s$ and $w_B=1$ for alleles $A$ and $B$, respectively, which affect
the probability $p_{\rm r}(x)$ that a new individual carries allele $A$ after reproduction as 
\be\label{prWF}
 p_{\rm r}(x)=\frac{w_A\Omega_A}{w_A\Omega_A+w_B\Omega_B}=\frac{(1+\tilde s)x}{1+\tilde s x}.
\ee
Alternatively, the dynamics of the same population can be described by the Moran model \cite{sMoran}.
At each time step of the dynamics two individuals (not necessarily distinct) are randomly selected in the population. In the absence of selection, an exact copy of the first one is introduced in order to replace the second one, which is therefore removed from the population.
Since individuals are randomly chosen, the probability $d_A=\Omega_A/\Omega=x$  of removing an individual with allele $A$  from the population equals the probability $r_A$ of reproducing one of them. Analogously, for an individual carrying allele $B$ these probabilities are $d_B=r_B=1-x$.
Within the Moran model, a selective advantage  can be accounted for by modifying the reproduction probability $r_{A,B}$ of the alleles with the fitnesses $w_A$ and $w_B$ specified above, according to $r_A(x)=(1+\tilde s)x/(1+\tilde s x)$ and 
$r_B(x)=x/(1+\tilde s x)$. With these probabilities, the number of individuals carrying allele $A$ increases/decreases by one at each step of the dynamics with rates $W_{+1}$/$W_{-1}$, respectively, with
\be
\label{rateMoran1}
 \begin{split}
  W_{+1}\delta t &=r_Ad_B=(1+\tilde s)x(1-x)/(1+\tilde s x),\\
  W_{-1}\delta t &=r_Bd_A=x(1-x)/(1+\tilde s x),
 \end{split}
\ee
where $\delta t$ is the duration of the time step \cite{notaMoran}.

Although the Wright-Fisher and Moran models are implemented with different rules at the microscopic level, for a wide range of values of the parameters and sufficiently large populations, they turn out to be effectively described by the same Langevin equation (with It\^o prescription, see Appendix \ref{app-diffusion})
\be
 \dot x=\mu(x)+\sqrt{v(x)}\,\eta(t),
 \label{eq:LG-DA}
\ee
where the evolution of the frequency $x$ of allele $A$ in the population is driven by the sum of a deterministic force $\mu(x)=\tilde s x(1-x)$ generated by selection and of a stochastic term --- referred to as \emph{genetic drift} in the literature --- which is a delta-correlated Gaussian noise with zero mean and variance $v(x)=x(1-x)/(\Omega\tau_g)$. 
This noise is conveniently expressed as $\sqrt{v(x)}\,\eta(t)$ in terms of the normalized Gaussian noise $\eta$ with $\langle \eta \rangle =0$ and $\langle\eta(t)\eta(t')\rangle=\delta(t-t')$.
Note that Eq.~\reff{eq:LG-DA} provides an approximate description of the dynamics in terms of an effective diffusion process. While this approximation turns out to be accurate for the Wright-Fisher and Moran models, at least within a suitable parameter range (see Refs.~\cite{prl,diffusion}), it is known to fail in other cases,  e.g., in the susceptible-infected-susceptible (SIS) model of epidemiology \cite{sis}.

Without loss of generality, time can be measured in units of generations, so that $\tau_g=1$ and the rates become dimensionless quantities. Balancing selection is characterized by a selective advantage $\tilde{s}$ which favors the evolution towards a state of the population characterized by an optimal frequency  $x_{\ast}$ of allele $A$; in the simplest case one can assume a linear dependence  $\tilde{s} = s(x_{\ast} - x)$, with a constant $s>0$. Note that in an infinitely large population (with $\Omega\to\infty$), the fluctuation effects represented by $\eta$ in Eq.~\reff{eq:LG-DA} are suppressed, and the resulting deterministic dynamics due to the selection term $\mu$  drives the population towards the optimal frequency $x_\ast$ of allele $A$. In a finite population, instead, the presence of fluctuations due to the random genetic drift eventually drives $x$ towards one of the two possible absorbing states $x=0$ and $1$, corresponding to the fixation of allele $B$ and $A$, respectively. 
\change{The mean fixation time and the fixation probability of a population described by Eq.~\reff{eq:LG-DA} can be evaluated within the diffusion approximation by the standard methods introduced in Ref.~\cite{kimura-ohta}. A summary of the relevant results and expressions is provided in the Appendix~\ref{app-mft}.}

In the absence of spatial embedding, a celebrated prototype model of subdivided populations is the so-called ``island model'', originally proposed by Wright \cite{wright} for neutral evolution. It consists of $N$ interacting subpopulations (\emph{demes}) of identical size $\Omega$, labeled by an integer $i=1,\ldots,N$ and characterized by the frequencies $\{x_1,x_2,\ldots ,x_N \}$ for the occurrence of allele $A$, with $x_i\in [0,1]$.
Within each deme, the internal dynamics (assumed to be identical in the absence of migration) proceeds as in either Moran's or Wright-Fisher's stochastic models, while different demes interact by exchanging randomly 
selected individuals, such that the sizes $\Omega$ of the demes involved in the exchange are not affected.
The rate $m$ with which migration occurs is defined as $m=n_{i\leftrightarrow j}N/\Omega$, where $n_{i\leftrightarrow j}$ is the mean number of individuals exchanged between the deme $i$ and $j$ in one generation.
\change{As a consequence of this exchange, the transition rates which define the Moran and Wright-Fisher models are modified as described in Appendix \ref{app-diffusion}.}
For sufficiently large $\Omega$ and small $m$ and $s$ \change{(see Appendix \ref{app-diffusion})}, the evolution of the allele frequency 
$x_i$ in the $i$-th deme can be described, both within the Moran and Wright-Fisher models by the following Langevin equation with It\^{o} prescription 
\be
\label{Langevin_single-deme}
\dot{x}_i =\mu(x_i)+ m (\bar x - x_i) +\sqrt{v(x_i)}\ \eta_i,
\ee
where $\eta_i$ are the independent Gaussian noises with correlation $\langle\eta_i(t)\eta_j(t')\rangle=\delta_{i,j}\delta(t-t')$, while the term $m(\bar x-x_i)$ accounts for the migration of individuals 
\change{between the demes \change{and it depends on $x_{j\neq i}$} only through the inter-deme mean frequency $\bar x=\sum_{i=1}^Nx_i/N$.}
%

In the presence of migration, each deme exchanges individuals with the others, a process that effectively acts as a source of biodiversity inside each deme, preventing them from achieving independent fixation. In fact, the single-deme states $x_i=0$ and 1 are no longer \emph{per se} absorbing for $m\neq 0$ and global fixation requires a coordinate evolution towards the two global absorbing states $X_0\equiv \{x_i=0\}_{i=1,\ldots,N}$ or 
$X_1\equiv \{x_i=1\}_{i=1,\ldots,N}$ in which all demes fixate the same allele.
In this case, the dynamics of the population can be conveniently described via the exact evolution equation for the mean frequency $\bar x$, which can be obtained directly from Eq.~\reff{Langevin_single-deme} 
\change{(see, e.g., Eq.~(S10) in the Supplemental Material of Ref.~\cite{prl}),} 
\begin{equation}
\label{Langevin_xbarra}
\dot{\avg{x}} =s[x_{\ast} \avg{x}-(1+x_{\ast})\avg{x^2} +\avg{x^3}] +
\sqrt{(\avg{x}-\avg{x^2})/(\Omega N)}\, \eta,
\end{equation}
where $\eta$ is a Gaussian noise with $\langle\eta(t)\eta(t')\rangle = \delta(t-t')$ and $\avg{x^k}=\sum_{i=1}^Nx_i^k/N$. 
Due to the non-linear nature of Eq.~\reff{Langevin_single-deme}, the evolution equation for $\bar x$ involves higher-order moments $\avg{x^2}$ and $\avg{x^3}$ which in principle could be determined by solving a whole hierarchy of coupled differential equations. 

\change{\subsection{Disentangling time scales}}
\label{sec:adiabatic}

\begin{figure}
 \centering
 \includegraphics[width=1\columnwidth]{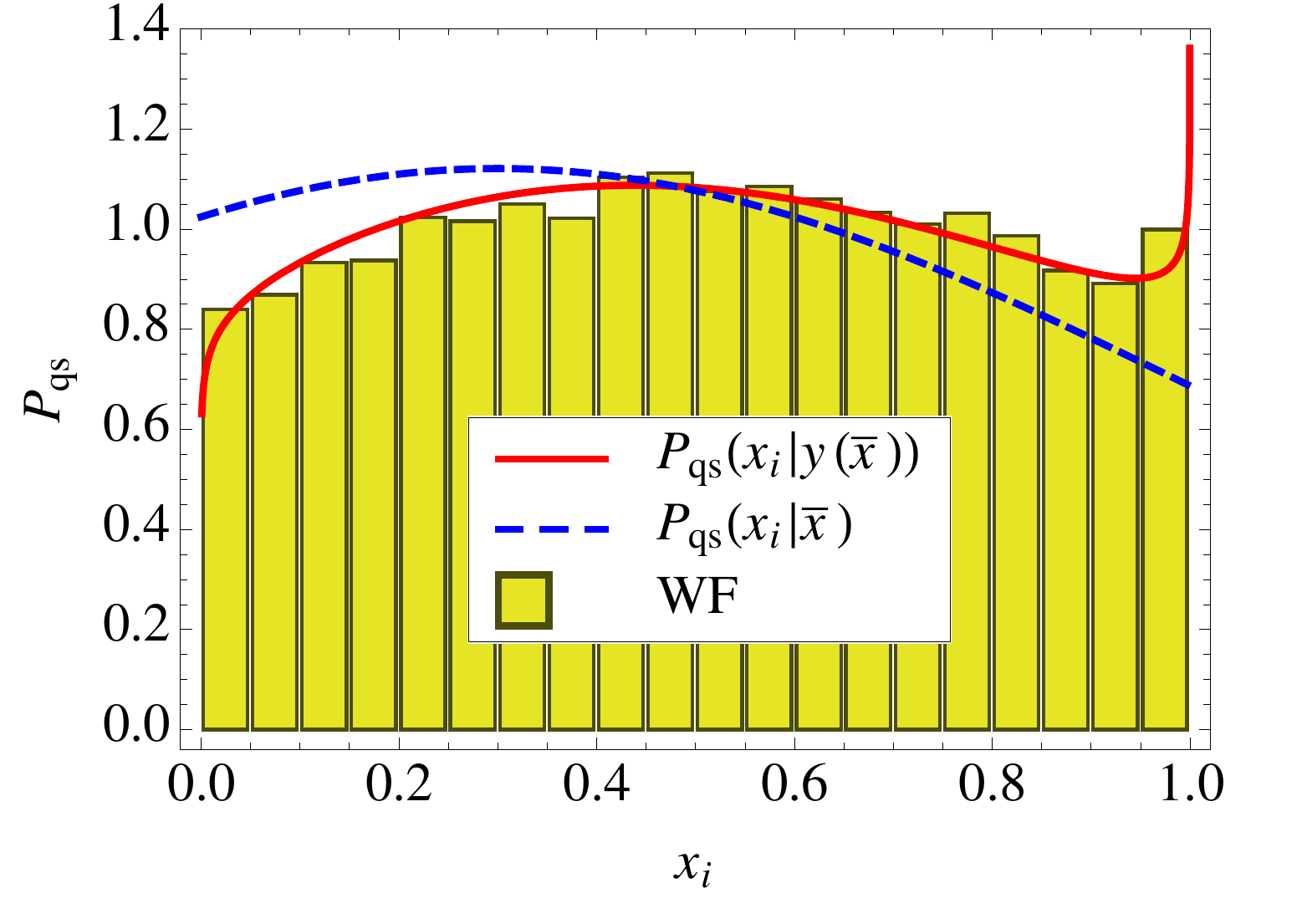}
 \caption{(Color online) 
Single-deme frequency distribution in a metapopulation of $N=100$ demes with $\Omega=100$ individuals each, with 
 $s'=m'=1$ and $x_*=0.3$, conditioned to having a mean frequency $\bar x = 0.5\pm 10^{-3}$. The histogram has been obtained by simulating the Wright-Fisher model (WF) \change{with migration (see Appendix~\ref{app-diffusion})}.
 The dashed line corresponds to the quasi-stationary distribution $P_{\rm qs}(x_i|\bar x = 0.5)$ in Eq.~\reff{eq:PqsFP}, while the solid line indicates $P_{\rm qs}(x_i|y(\bar x = 0.5))$ with $y(\bar x=0.5) \simeq 0.53$ determined by solving numerically the consistency equation~\reff{self-cons}.
 \change{The histogram refers to the set of single-deme frequencies $\{x_i\}$ recorded during the evolution of the population after an initial transient but much before fixation occurs and such that the corresponding fluctuating mean $\bar x(t)$ is close to 0.5, i.e., with $|\bar x(t)-0.5| < 10^{-3}$. In order to avoid correlations between successive recordings, we assumed a minimal time-lag of $10\,T_{\rm corr}$ where $T_{\rm corr} =1/m$ is the time scale associated with a change of $x_i$ of the same order as $x_i$, due to migration.}
 }
  \label{fig:distribution}
\end{figure}

As recently discussed in Ref.~\cite{prl}, for a large number of demes $N$ and small selection strength $s$
\change{(see \cite{notatime scale} for a precise statements of these conditions),} 
 a time scale separation emerges between the local dynamics and the global one, i.e., between the dynamics of $x_i$ and that of $\bar x$: \change{this
allows one to express the moments $\avg{x^k}$ in Eq.~\reff{Langevin_xbarra} in terms of $\bar x$.}
In fact, on the typical time scale of variation of the inter-deme mean frequency $\bar x$, the local frequencies $\{x_i\}_{i=1,\ldots, N}$ are rapidly fluctuating and their statistics can be described by a quasi-stationary distribution $P_{\rm qs}(x_i|\bar x)$. Conversely, on the time scale which characterizes the fast fluctuations of each single local frequency $x_i$, the mean frequency $\bar x$  is practically constant. 
\change{In passing, we mention that an analogous separation of time scales occurs in non-homogeneous metapopulations~\cite{KKS-12}.
In the present case, it} 
suggests that the quasi-stationary distribution could be accurately approximated by the stationary solution of the Fokker-Planck equation associated with Eq.~\reff{Langevin_single-deme} with fixed $\bar{x}$, which is 
\be 
P_{\rm qs}(x_i|\bar x) \propto x^{2m'\bar x-1}(1-x)^{2m'(1-\bar x)-1}{\rm e}^{s'x(2x_*-x)},
\label{eq:PqsFP0}
\ee
where $m'=\Omega m$ and $s'=\Omega s$ are a conveniently rescaled migration rate and selection coefficient. 
\change{Figure \ref{fig:distribution} shows the histogram of the single-deme frequencies $x_i$ in a certain 
metapopulation with a mean $\bar x =0.5$, as obtained from the numerical simulation of the Wright-Fisher model. 
%
The dashed line in the figure indicates the theoretical prediction given by the quasi-stationary distribution $P_{\rm qs}(x_i|\bar x = 0.5)$ in Eq.~\reff{eq:PqsFP} and it clearly shows significant discrepancies with the actual histogram. 
According to Ref.~\cite{prl}, this discrepancy can be resolved by considering a quasi-stationary distribution of the same functional form as Eq.~\reff{eq:PqsFP0} but in which an effective parameter $y(\bar x)$ replaces $\bar x$ in order to account for the fact that the latter actually varies in time:
\be
P_{\rm qs}(x|y)\propto x^{2m'y-1}(1-x)^{2m'(1-y)-1}{\rm e}^{s'x(2x_*-x)}.
\label{eq:PqsFP}
\ee
In turn, $y(\bar x)$ is determined in such a way to satisfy the consistency condition 
\be\label{self-cons}
\bar x=\int_0^1\rmd x\,x\,P_{\rm qs}(x|y(\bar x)).
\ee}
\change{When the selection coefficient $s$ vanishes, the time scale separation is very pronounced and this equation gives $y(\bar x)=\bar x$. 
In the presence of a non-vanishing selection strength, one can either solve Eq.~\reff{self-cons} numerically or do an expansion} in the small parameter $s_{\rm e}/m$, with 
\be\label{se} 
 s_{\rm e}=\frac{s}{\left(1+\frac{1}{m'}\right)\left(1+\frac{1}{2m'}\right)},
\ee
which renders
\be
y(\bar x)=\bar x-(s_{\rm e}/m)\bar x(1-\bar x)(x_*^{\rm e}-\bar x)+O((s_{\rm e}/m)^2),
\ee
where  
\be \label{xe} 
x_*^{\rm e}=x_*+(x_*-1/2)/m'
\ee 
is another effective parameter that will be discussed further below. 
\change{By solving Eq.~\reff{self-cons} for the choice of parameters in Fig.~\ref{fig:distribution}, one obtains $y(\bar x = 0.5) \simeq 0.53$;  the corresponding distribution $P_{\rm qs}(x_i|y(\bar x = 0.5))$ is reported as a solid line in Fig.~\ref{fig:distribution} and, in fact, it turns out to describe the actual distribution significantly better than $P_{\rm qs}(x_i|\bar x = 0.5)$. We emphasize the fact that in this comparison there are no fitting parameters. A similar improvement is found also for different values of $\bar x$ and of the parameters which characterize the population.}

%
%
%
%
%

%
When the time scales separation holds, the moments $\avg{x^k}$ on the r.h.s.~of Eq.~\reff{Langevin_xbarra} can be approximated by the corresponding moments $\langle x^k\rangle=\int_0^1 \rmd x\, x^kP_{\rm qs}(x|y(\bar x))$, leading to the effective Langevin equation
 \be\label{Langevin_xbarra_gen}
  \dot{\bar  x}=M(\bar x)+\sqrt{V(\bar x)}\,\eta(t),
 \ee
where the deterministic term and the variance of the noise are respectively given by 
\begin{subequations} 
 \begin{align}
 &M(\bar x)=s\int_0^1\rmd x\, x(1-x)(x_*-x)P_{\rm qs}(x|y(\bar x)) \label{eq:Meff}\\
 \mbox{and}\nonumber\\
&V(\bar x)=(\Omega N)^{-1}\int_0^1 {\rm d}x\,x(1-x)P_{\rm qs}(x|y(\bar x)).
 \end{align}
\end{subequations}
At the lowest non-trivial order in small $s_{\rm e}/m$, the population behaves as a well-mixed one with effective 
parameters which are rescaled due to the finite migration rate $m$, in agreement with the results of Refs.~\cite{maruyama,slatkin,cherry-wakeley}.
In particular, the deterministic force and the variance of the stochastic term turn out to be 
\begin{subequations} 
\label{eq:MV0}
\begin{align}
M^{(0)}(\bar x) & =s_{\rm e}\bar x(1-\bar x)(x_*^{\rm e}-\bar x), \label{m0x}\\
\label{v0x} V^{(0)}(\bar x) & =\bar x(1-\bar x)/N_{\rm e},
\end{align}
\end{subequations} 
where the {\em effective selection strength} $s_{\rm e}$ and the {\em effective optimal frequency} $x_*^{\rm e}$ are given in Eq.~\reff{se} and \reff{xe} respectively, while  
\be \label{ne}
N_{\rm e}=N\Omega\left(1+\frac{1}{2m'}\right)
\ee 
is the {\em effective population size}.  
\change{(In Eqs.~\reff{eq:MV0} and in what follows, the superscript $(0)$ indicates that the corresponding quantity has been calculated at the lowest order in an expansion in $s_{\rm e}/m$.)}
It is worth noting that as the migration rate $m'$ increases, the effective parameters  $s_{\rm e}$ and $x_{\rm e}^*$ approach the values they have for the isolated demes, while the effective size $N_{\rm e}$ tends to the total number $N\Omega$ of individuals in the metapopulation;
accordingly, in the limit $m'\to\infty$, the internal structure of the metapopulation does not affect its dynamics and subdivision plays no actual role.
\change{(This might not be the case in non-homogeneous populations, as discussed  in Ref.~\cite{KKS-12}.)}

If balancing selection is not symmetric, i.e., $x_*\neq 1/2$, the effective drift $M^{(0)}(\bar x)$ in Eq.~\reff{m0x} can be written as the sum of a symmetric term 
\be
M^{(0)}_{\rm symm}(\bar x)=s_{\rm e}\bar x(1-\bar x)(1/2-\bar x)
\ee
 and a directional selection term \cite{nota dirsel}
 \be\label{Mdir}
 M^{(0)}_{\rm dir}(\bar x)=\sigma_{\rm e}\bar x(1-\bar x),
 \ee
where 
\be\label{sigmae}
\sigma_{\rm e}=s_{\rm e}(x_*^{\rm e}-1/2)
\ee
is an \emph{effective directional selection coefficient}.
$M^{(0)}_{\rm symm}$ promotes coexistence of the two alleles and therefore it increases the biodiversity of the system, slowing down fixation;  $M^{(0)}_{\rm dir}$, instead, favors fixation of one of the alleles (depending on the sign of $\sigma_{\rm e}$). The competition between these two terms determines whether balancing selection actually slows down or speeds up fixation of the population as a whole. One can therefore expect that, depending on the ratio $|\sigma_{\rm e}/s_{\rm e}|$ being larger than some threshold $\theta$, 
$M^{(0)}_{\rm dir}$ prevails over $M^{(0)}_{\rm symm}$ such that balancing selection eventually accelerates fixation.
According to Eq.~\reff{xe}, this occurs for 
\be
\label{PT_heuristic}
 |x_*-1/2|>\frac{m'\theta}{m'+1},
\ee
which provides a heuristic estimate of the region of the parameter space within which balancing selection should facilitate fixation. 
The fact that balancing selection slows down fixation only if the optimal frequency $x_*$ is far enough from the absorbing boundaries (i.e., if it is close enough to $x_*=1/2$) was first noticed in Ref.~\cite{robertson} for the case of balancing selection in well-mixed populations, by analyzing the eigenvalues of the transition matrix of the Moran-like dynamics.
In the next section, we argue that this change of behavior becomes an actual phase transition in the limit $N \to \infty$ of subdivided populations.

\section{Phase transition in the infinite island model ($N=\infty$)}
\label{sec:PT}

\begin{figure}[t] 
\includegraphics[width=\columnwidth]{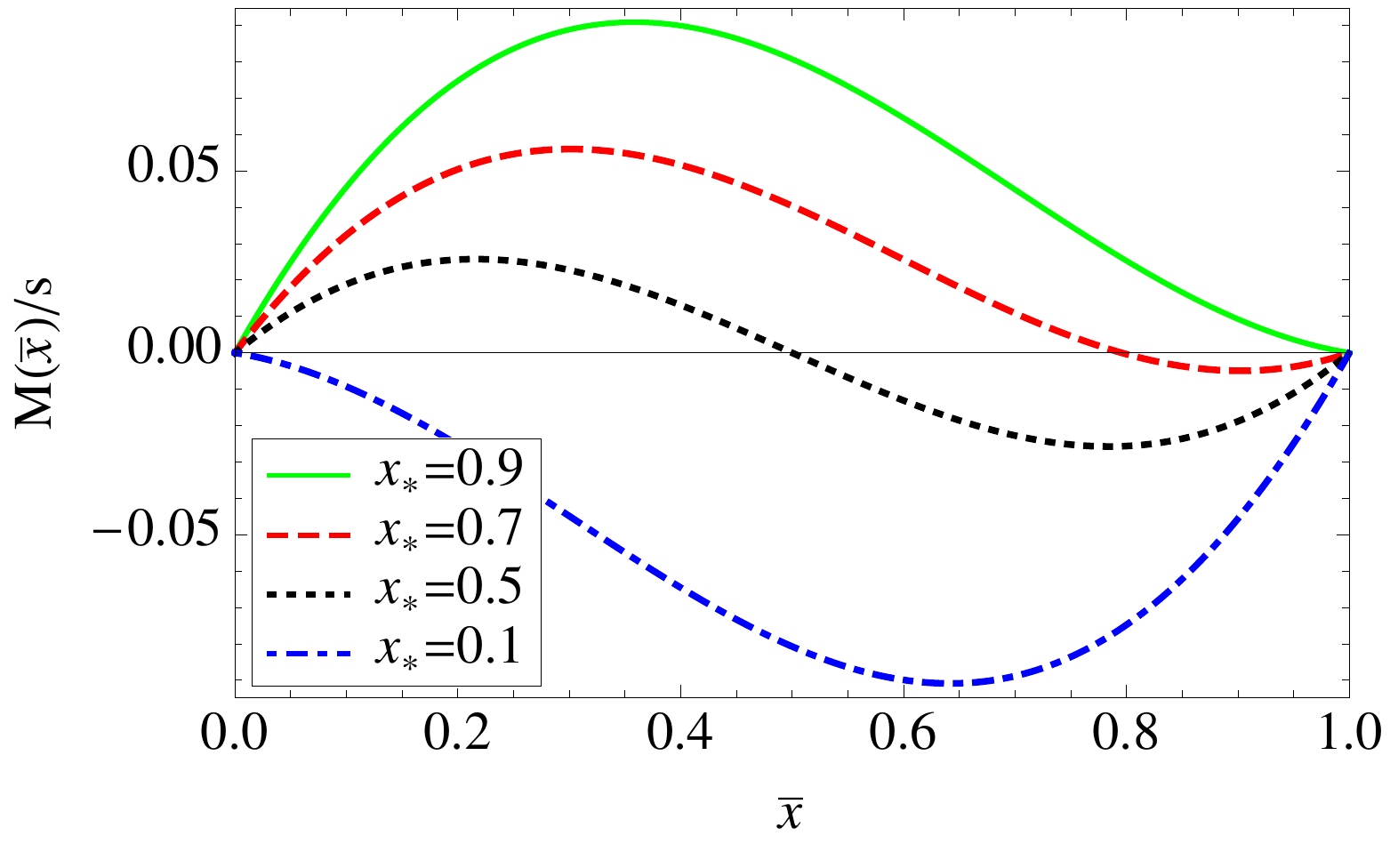}
\caption{(Color online) Drift $M(\bar x)$ as a function of $\bar x$, as obtained from the numerical solution of Eq.~\reff{eq:Meff} 
for $m'=2$, $s'=1$
\change{(corresponding to $s_{\rm e}/m\simeq 0.3$)}, 
and for various values of $x_*$. It can be noticed that, depending on the value of $x_*$, a non-trivial zero $x_{\infty}$ emerges, which is always an attractive state for the deterministic evolution of $\bar x$.}
\label{fig:M}
\end{figure}

\begin{figure}
 \centering
 \includegraphics[width=1\columnwidth]{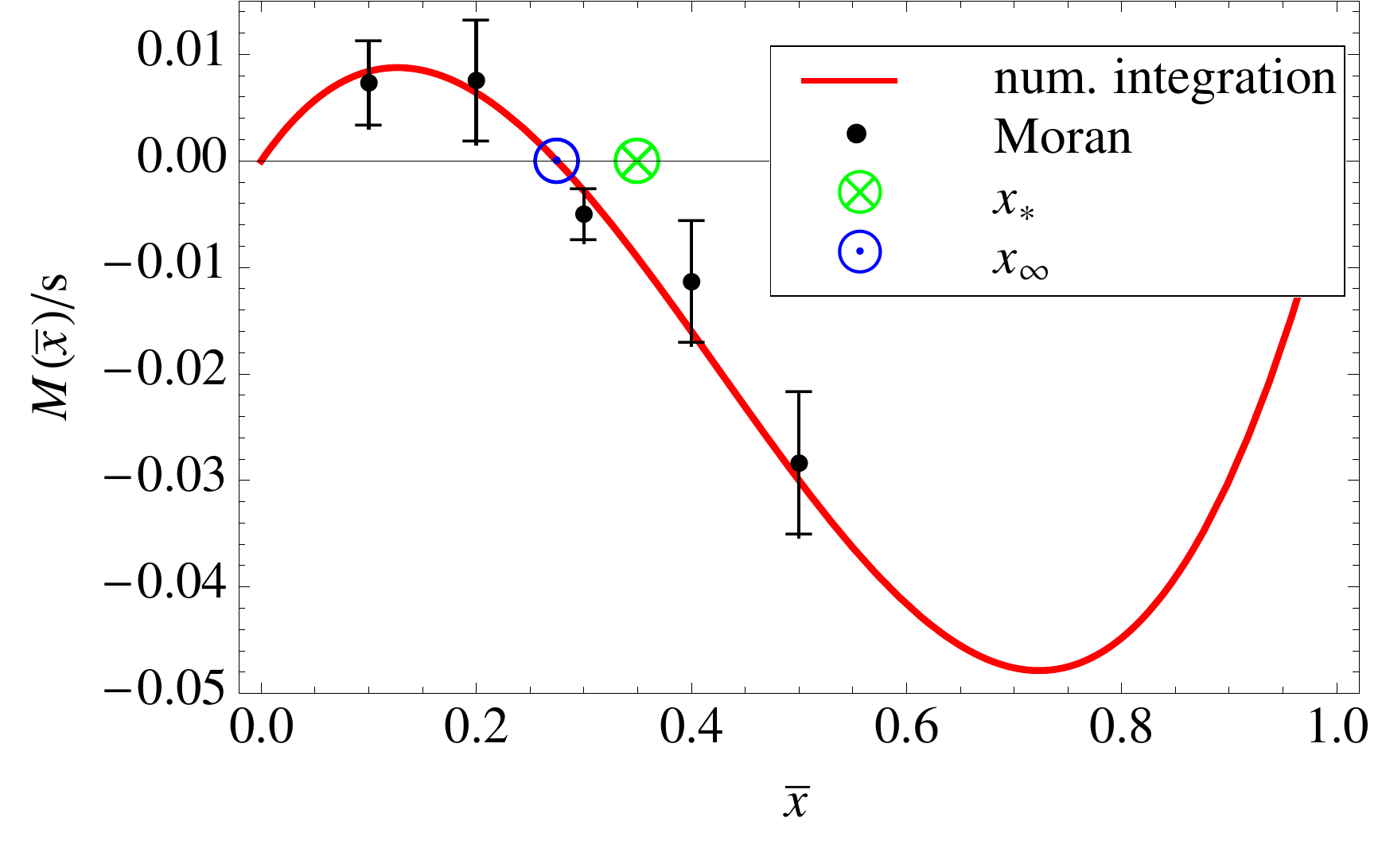}
 \caption{(Color online)  
 \change{Deterministic term $M(\bar x)$ as a function of the mean frequency $\bar x$ in a population of $N=100$ demes with $\Omega=100$ individuals each and parameters $m'=2$, $s'=1$, and $x_*=0.35$ (this latter value is indicated by the crossed circle). The solid line corresponds to the numerical solution of Eq.~\reff{eq:Meff}, from which we can read the estimate of $x_\infty$ (empty circle). Symbols with errorbars, instead, are the results of the  numerical simulations of this metapopulation, based on the Moran model (see Appendix~\ref{app-diffusion} for the definition of the rates).
In order to estimate $M(x_0)$ from the numerical data, the increment $\delta \bar x(t)$ is recorded at each Moran step of the dynamics such that, correspondingly,  $|\bar x(t)-x_0|<10^{-3}$, where $\bar x(t)$ is the fluctuating mean value of $x_i(t)$ within the population. These increments are recorded from a time $5\,T_{\rm corr}$ 
after the beginning of the evolution to well before the eventual fixation occurs, and their mean gives $M(x_0)$. This figure can be compared with Fig.~\ref{fig:M} corresponding to different values of $x_*$.}
 }
  \label{fig:Mxm}
\end{figure}

In the limit of an infinite number of demes, Eq.~\reff{Langevin_xbarra_gen} becomes deterministic because the variance $V(\bar x)$ of the noise vanishes; this is not the case for the noise in the single-deme equation~\reff{Langevin_single-deme}, which is finite as long as  $\Omega$ is finite and therefore determines a non-trivial quasi-stationary distribution $P_{\rm qs}(x|y(\bar x))$ which, in turn, affects Eq.~\reff{Langevin_xbarra_gen}.
Depending on the values of the parameters $s'$, $m'$, and $x_*$, the internal stochasticity of the demes might be sufficiently strong  to drive the metapopulation to fixation even in the infinite-size limit $N\to\infty$. In fact, the deterministic part of Eq.~\reff{Langevin_xbarra_gen} might drive $\bar x$ towards one of the two absorbing states $X_0$ and $X_1$ corresponding to $\bar x=0$ and $\bar x =1$, respectively. 
In addition to these latter solutions, Eq.~\reff{Langevin_xbarra_gen} admits also a stationary state with $\bar x=x_{\infty}$, which can be determined by requiring that $M(x_\infty)$ vanishes, i.e., by solving the equation [see Eq.~\reff{eq:Meff}]
\be
\label{metastable_eq}
   \int_0^1\rmd x\, x(1-x)(x_*-x)P_{\rm qs}(x|y)=0,
\ee
where $y=y(x_{\infty})$ is defined by the consistency condition in Eq.~\reff{self-cons}.
Figure~\ref{fig:M} shows the numerical determination of $M(\bar x)$ 
\change{(based on Eqs.~\reff{self-cons} and \reff{eq:Meff})} 
as a function of $\bar x$ for various values of $x_*$ in a population characterized by the parameters reported in the caption. 
%
%
%
%
%
%
\change{Figure~\ref{fig:Mxm}, instead, shows the comparison between $M(\bar x)/s$ calculated as in Fig.~\ref{fig:M} (solid line) for $x_*=0.35$ (indicated by the crossed circle) and the one inferred from the numerical simulations (symbols with errorbars) of a population with $N=100$ demes of $\Omega=100$ individuals each and the same values of parameters as in Fig.~\ref{fig:M}. The evolution was performed according to the Moran model, as the latter turns out to be numerically more efficient for the determination of $M$ than the Wright-Fisher model considered in Fig.~\ref{fig:distribution}.
This comparison shows that the effective description introduced in Sec.~\ref{sec:adiabatic} captures also the quantitative aspects of the actual dynamics of the subdivided population, at least within the range of parameters investigated here. In particular, $x_\infty$ computed from Eq.~\reff{metastable_eq} provides an accurate estimate of the one inferred from numerical simulations (circle in the figure).
}
%
%
%
%
%
%
%
\change{This} non-trivial zero  $x_{\infty}$ of $M$ \change{in Figs.~\ref{fig:M} and \ref{fig:Mxm}} corresponds to an attracting stable state for the deterministic part of Eq.~\reff{Langevin_xbarra_gen},  which is asymptotically reached for $t\to\infty$ unless the initial conditions are exactly on a boundary.  The stability of the point $x_\infty$ follows from the fact that $M'(x_\infty) < 0$.
When $x_{\infty}\in(0,1)$, it represents a stable ``active'' state for the infinite population and it corresponds to the infinite-size limit ($N\to\infty$) of the metastable state in which a finite system ($N<\infty$) would spend a long time before reaching fixation \cite{prl}.
However  $x_{\infty}$  might coincide with one of the two boundaries 0 and 1, depending on the values of the parameters $x_*$, $m'$, and $s'$ and correspondingly $M(\bar x)$ has the same sign within the whole interval $(0,1)$: when this happens, the deterministic part of the dynamics drives the system towards fixation.
Note that this fixation process is deterministic in nature and the system always reaches (asymptotically in time) the absorbing state determined by $x_{\infty}$, differently from the case with finite $N$ in which fixation is a stochastic process and both boundaries are attainable.

\begin{figure}[t] 
\includegraphics[width=1.1\columnwidth]{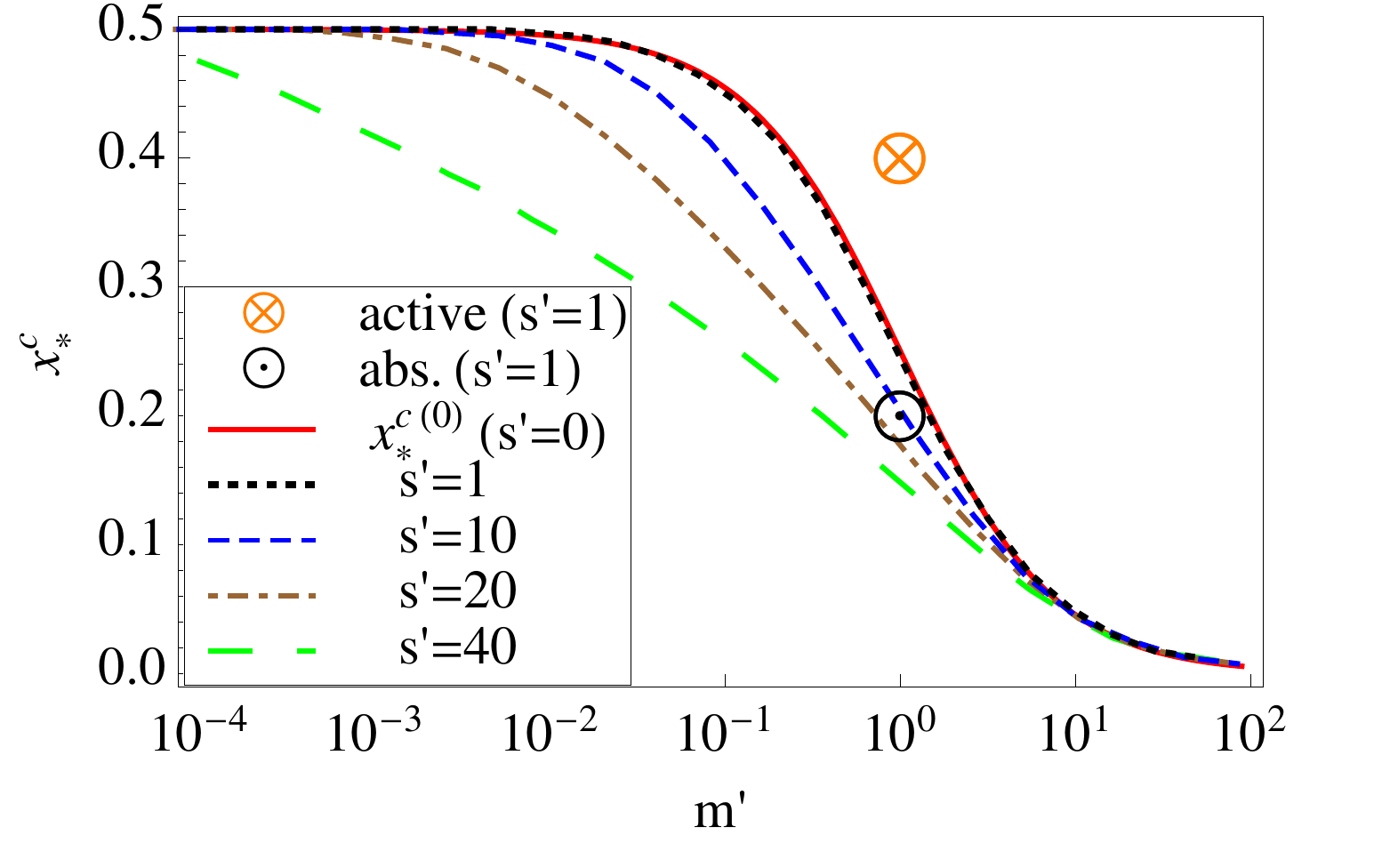}
\caption{(Color online) Critical value $x_*^{\rm c}$ as a function of $m'$ for several values of $s'$. For $x_*< x_*^{\rm c}$ or $x_*>1-x_*^{\rm c}$ the system is driven deterministically to fixation for $N\to\infty$.
The solid line is the  estimate in Eq.~\reff{xcritico_stima}, valid for small $s$, while the dashed lines correspond to the numerical solution of Eq.~\reff{metastable_eq} for larger values of $s'$. 
\change{Focussing on the case $s'=1$ and $m'=1$, metapopulations with $N\to\infty$ and $x_*$ larger than $x_*^c \simeq 0.25$ (dashed curve) are expected to be in the active phase, whereas those with $x_*$ smaller than that rapidly fixate. This expectation is confirmed by numerical simulations of metapopulations with finite but large $N$: the crossed and dotted circles indicate the values of $x_*$ for which there is numerical evidence for them to correspond to an active and an absorbing phase, respectively, as discussed in, c.f., Fig.~\ref{fig:scaling} and further below.}
}
\label{fig:PT}
\end{figure}

When computing the stationary value $x_{\infty}$ as a function of the optimal frequency $x_\ast$ for fixed $s'$ and $m'$, there exists a critical value $x_*^{\rm c}(s',m')$, such that for $x_*\in(x_*^{\rm c},1-x_*^{\rm c})$ the infinite population is in the active phase, i.e., $x_{\infty}\in(0,1)$, while it otherwise reaches one of the two absorbing states $x_{\infty} = 0$ or $1$. 
Figure~\ref{fig:PT} displays the dependence of the critical value $x_{\ast}^c$ on the migration rate $m'$ for several values of selection strength $s'$.
\change{In addition, for $m'=1$ and $s'=1$, Fig.~\ref{fig:PT} compares the prediction of having fixation for $x_*<x_{\ast}^c\simeq 0.25$ and an active state for  $x_*>x_{\ast}^c\simeq 0.25$ with the numerical evidences discussed further below (see also, c.f., Fig.~\ref{fig:PT}), corresponding to the conditions indicated by the dotted and crossed circles, respectively.}
Analytic estimates for the stationary value $x_{\infty}$ and for the critical value $x_*^{\rm c}$ can be easily obtained for small $s_{\rm e}/m$, in which case the condition \eqref{metastable_eq}  reduces to $M^{(0)}(\bar x)=0$. Using Eq.~\eqref{m0x} one gets 
\be
x_{\infty}^{(0)} = 
\left\{\begin{array}{cl}
  x_*^{\rm e} &{\rm for}\ x_*^{\rm e}\in[0,1],\\
  0&{\rm for}\ x_*^{\rm e}<0,\\
  1 &{\rm for}\ x_*^{\rm e}>1,
\end{array}
\right.
\ee
where $x_*^{\rm e}$ is given in Eq.~\reff{xe}, while
\be\label{xcritico_stima} 
x_*^{{\rm c}(0)} = \frac{1}{2(m'+1)}.
\ee
\change{(We remind here that the superscript $(0)$ denotes that the corresponding quantity has been calculated on the basis of the zeroth-order approximation $M^{(0)}(\bar x)$ for $M(\bar x)$ in Eq.~\reff{Langevin_xbarra_gen}.)}  
This expression agrees with the heuristic estimate of Eq.~\eqref{PT_heuristic} if one sets the numerical threshold $\theta$ to $\theta =1/2$.
The analytic determination of $x_*^{{\rm c}(0)}$ in Eq.~\eqref{xcritico_stima} is reported in Fig.~\ref{fig:PT} as a solid line and it coincides, as expected, with the estimate based on the numerical solution of Eq.~\reff{metastable_eq} for small $s'$ (uppermost dashed line).
Within the same approximation, the mean frequency $\bar x$ in the active phase approaches the effective optimal frequency $x_*^{\rm e}$ exponentially fast in time, i.e., $\bar x(t)-x_*^{\rm e}\propto\exp[-s_{\rm e}x_*^{\rm e}(1-x_*^{\rm e})t]$. In the absorbing phase, instead, an equally rapid evolution drives the system to fixation: for example $\bar x(t)\propto{\rm exp}[-s_{\rm e}|x_*^{\rm e}| t]$ for $x_*^{\rm e}<0$, with an equivalent expression holding for $x_*^{\rm e}>1$.
%
%
%
%

\change{According to Fig.~\ref{fig:PT}, a population with a certain $m'$ and $x_*$ may undergo the transition between coexistence and fixation upon varying $s'$, while for $x^* = 1/2$ coexistence is maintained for any positive value of $s'$.  
In this respect, the present phase transition resembles the one observed numerically in Ref.~\cite{lavrentovich} for 
the dynamics of the stepping-stone model \cite{stepping-stone} in two spatial dimensions with mutualistic forces, which has been verified experimentally in bacterial populations with similar properties \cite{momeni,muller}.
In addition, in the case of populations embedded in one spatial dimension, this fixation-coexistence phase transition --- emerging in the limit of infinite size --- has been argued \cite{korolev1d,dallasta1d} to belong to the so-called DP2 universality class \cite{hinrichsen}, with which it shares only some universal features (but, generically, not quantities such as the MFT). 
The analytical results presented here are in fact in agreement with the behavior expected for the DP2 phase transition within the mean-field approximation.}

\begin{figure}[t] 
\includegraphics[width=\columnwidth]{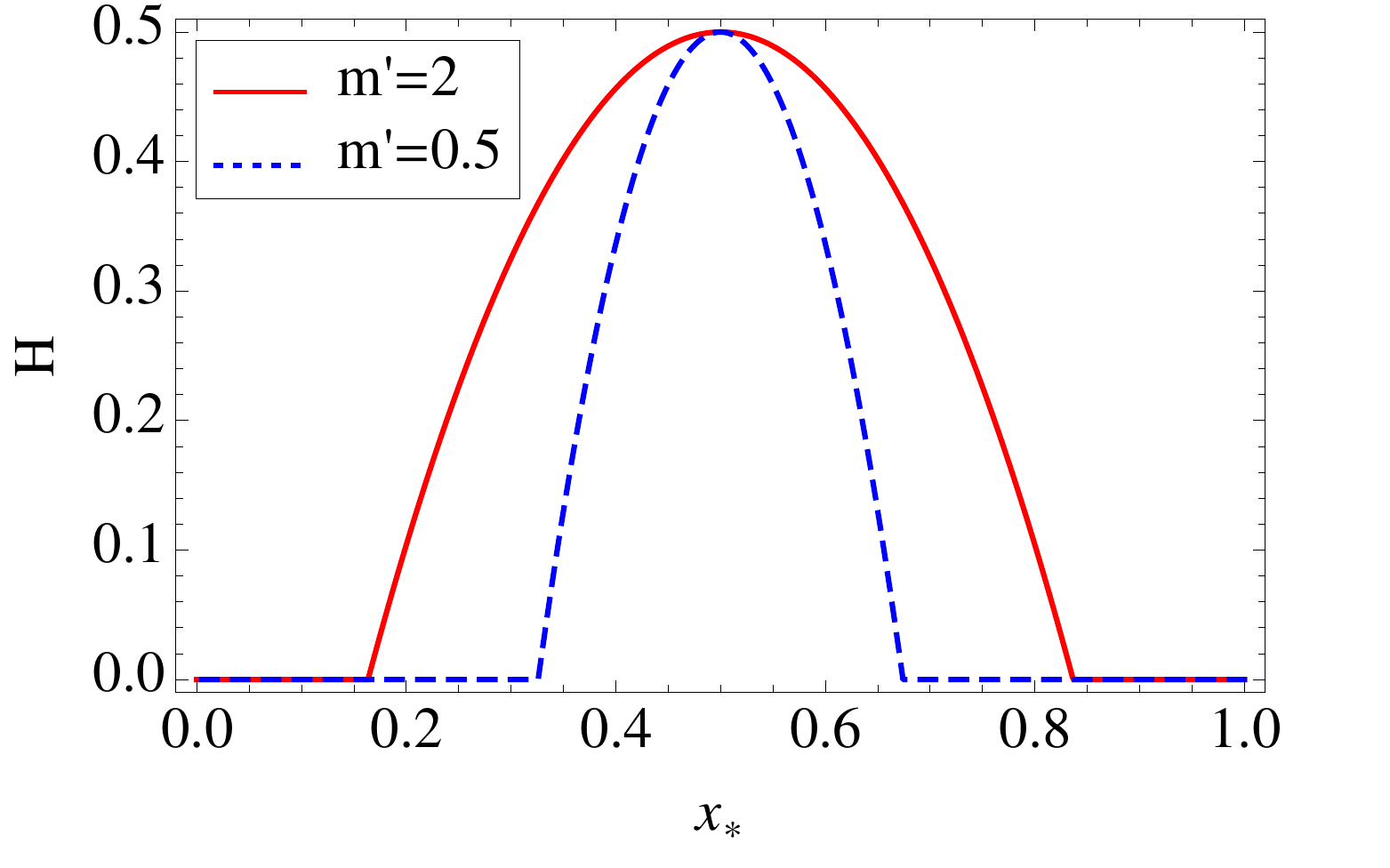}
\caption{(Color online) Global heterozygosity $H$ in the stationary state $\bar x=x_{\infty}$ for $N\to\infty$, 
as a function of the optimal frequency $x_*$ for $s'=1$ and various values of $m'$. 
The value of $H$ has been obtained by the numerical solution of Eq.~\reff{metastable_eq}.
}
\label{fig:exponent}
\end{figure}

The global heterozygosity $H=2\bar x(1-\bar x)$ provides an index of the biodiversity of the population, as it vanishes in the absorbing states $X_0$ and $X_1$, while it does not in the active phase. In this respect it can be considered as an order parameter for the phase transition occurring at $x_*= x_*^{\rm c}$ and  $x_*= 1-x_*^{\rm c}$. Figure~\ref{fig:exponent} reports $H$ as a function of $x_*$, as obtained from the numerical solution of Eq.~\reff{metastable_eq} for $s'=1$ and $m'=1$.
%
%
%
%

\begin{figure}[h!]
 \centering
 \includegraphics[width=1\columnwidth]{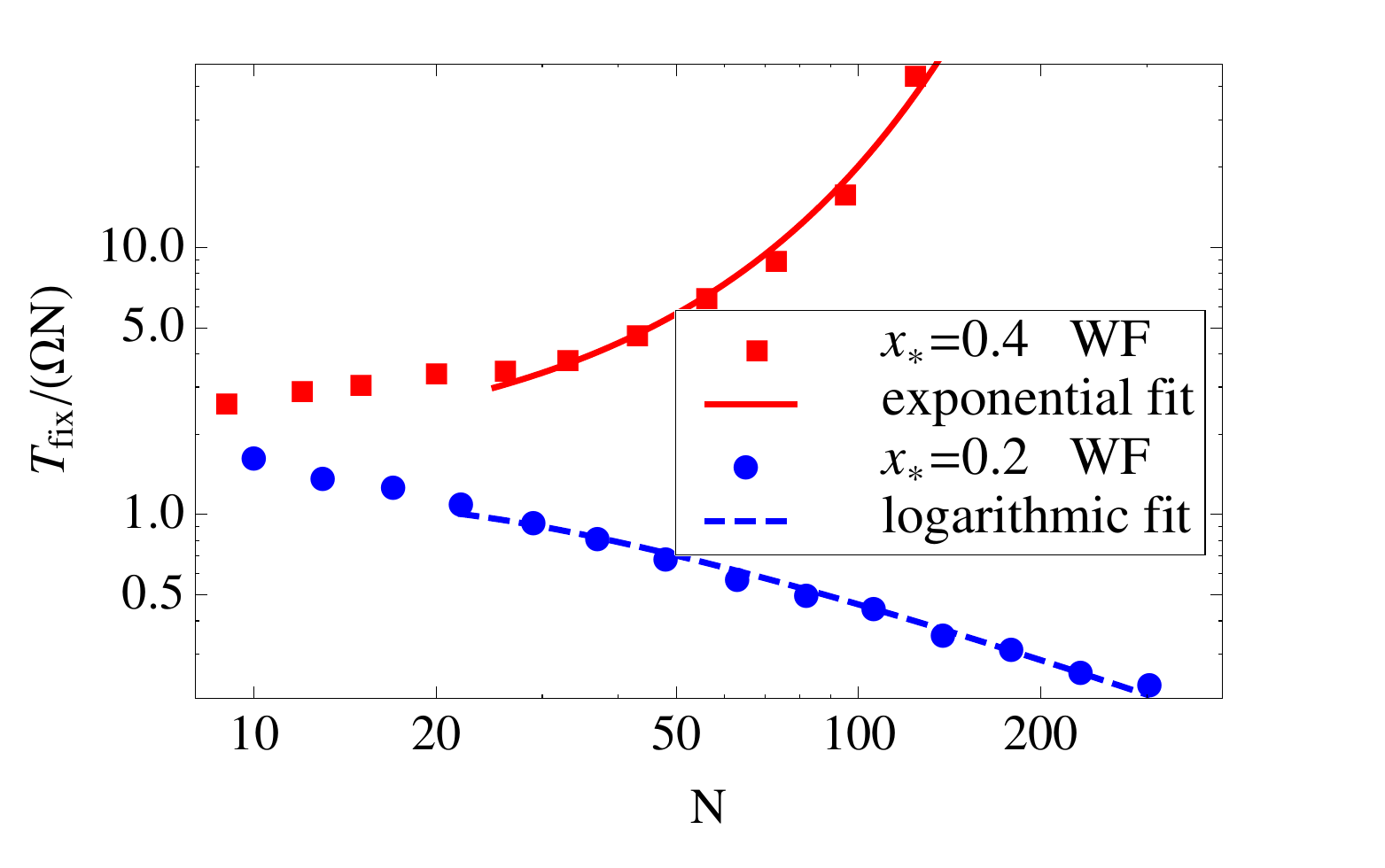}
 \caption{(Color online) Dependence of the MFT $T_{\rm fix}$ on the size $N$ of the metapopulation, for $\Omega=100$, $s'=1$ and $m'=1$. The MFT has been determined via numerical simulations of the Wright-Fisher (WF) model. For $x_*=0.4$ (red squares) $T_{\rm fix}/N$ shows an approximate exponential increase upon increasing $N$ (red solid line, $T_{\rm fix}/(\Omega N)=\exp\left[0.46+0.025N\right]$) while for $x_*=0.2$ (blue circles) $T_{\rm fix}$ displays an approximate logarithmic dependence on $N$ (blue dashed line, $T_{\rm fix}/(\Omega N)=\left[-27+16\,{\rm log}N\right]/N$).
 }
  \label{fig:scaling}
\end{figure}

The picture presented above approximately carries over to the case in which the number of demes $N$ is finite but large. In this case one can heuristically assume that, in the absorbing phase $x_* < x_*^c$ or $x_*>1-x_*^c$, fixation to a boundary ($\bar x =0$ or 1) is effectively reached when the distance of $\bar x$ to that boundary is smaller than $1/(\Omega N)$ (corresponding to having in the metapopulation only one individual different from the others) and therefore we expect the MFT to scale as $T_{\rm fix}\propto \log (\Omega N)$ because of the exponential law with which $\bar x(t)$ approaches the boundary as a function of time.
Figure~\ref{fig:scaling} reports the MFT as a function of $N$ for various values of the optimal frequency $x_*$. 
It can be noticed that, as expected from the arguments presented above, 
$T_{\rm fix}/N$ increases upon increasing $N$ in the active phase (red squares), while it decreases in the absorbing phase (blue circles),
and this supports the fact that a \emph{bona-fide} phase transition should be present in the limit $N\to\infty$. 
A numerical interpolation reveals indeed an exponential dependence of $T_{\rm fix}/N$ (red solid line) as a function of $N$ in the active phase while a logarithmic one (blue dashed line) of $T_{\rm fix}$ in the absorbing phase.

%
%
%
%

\section{The case of slow migration}
\label{sec:small_m}

The effective description of the island model with migration introduced in Ref.~\cite{prl} and summarized in Sec.~\ref{sec:adiabatic} relies on a perturbative expansion in the parameter $s_{\rm e}/m$. The predictions concerning the collective behavior of the metapopulation and, in particular, the mean fixation time are therefore valid only within the region of the parameter space corresponding to small $s_{\rm e}/m$. If the migration rate $m$ is large, this region stretches and includes large values of the selection strength $s\simeq m$. 
Interestingly enough, this case can also be described by using a fast-mode elimination method recently proposed in Ref.~\cite{fastmodes}. For small migration rate $m$, the approximation discussed in Sec.~\ref{sec:adiabatic} is expected to be accurate only for a small selection coefficient $s$. In particular, it requires $s\ll1/\Omega$ for the value $m\simeq 1/(\sqrt{2}\Omega)$ of the migration rate $m$ at which the parameter $s_{\rm e}/m$ reaches its maximum as a function of $m$.
\change{However, Fig.~2 in Ref.~\cite{prl} 
suggests that the approximation discussed in Sec.~\ref{sec:adiabatic} provides accurate predictions beyond the cases mentioned above.} 
In order to rationalize this fact, in this Section we develop an alternative description of the system for small migration rate $m$.

\subsection{Effective voter model}
\label{sec:voter}

\begin{figure}[h!]
 \centering
 \includegraphics[width=1\columnwidth]{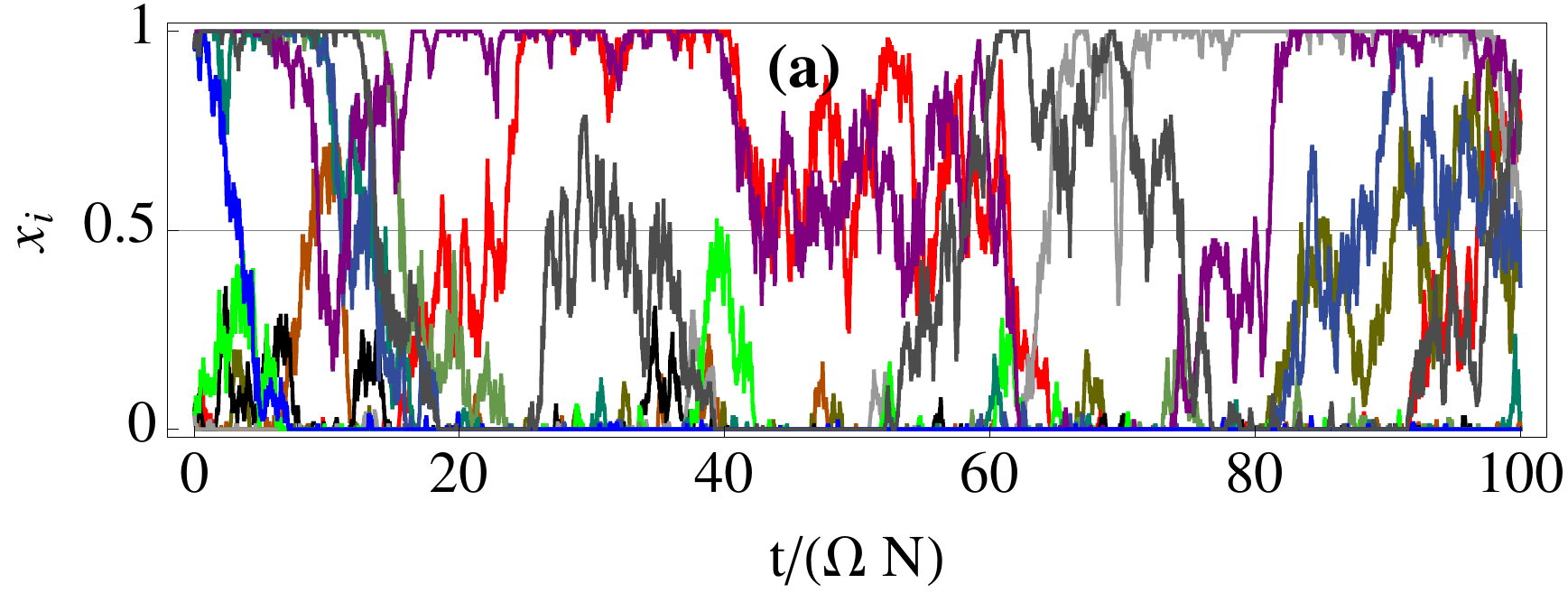}\\ 
  \includegraphics[width=1\columnwidth]{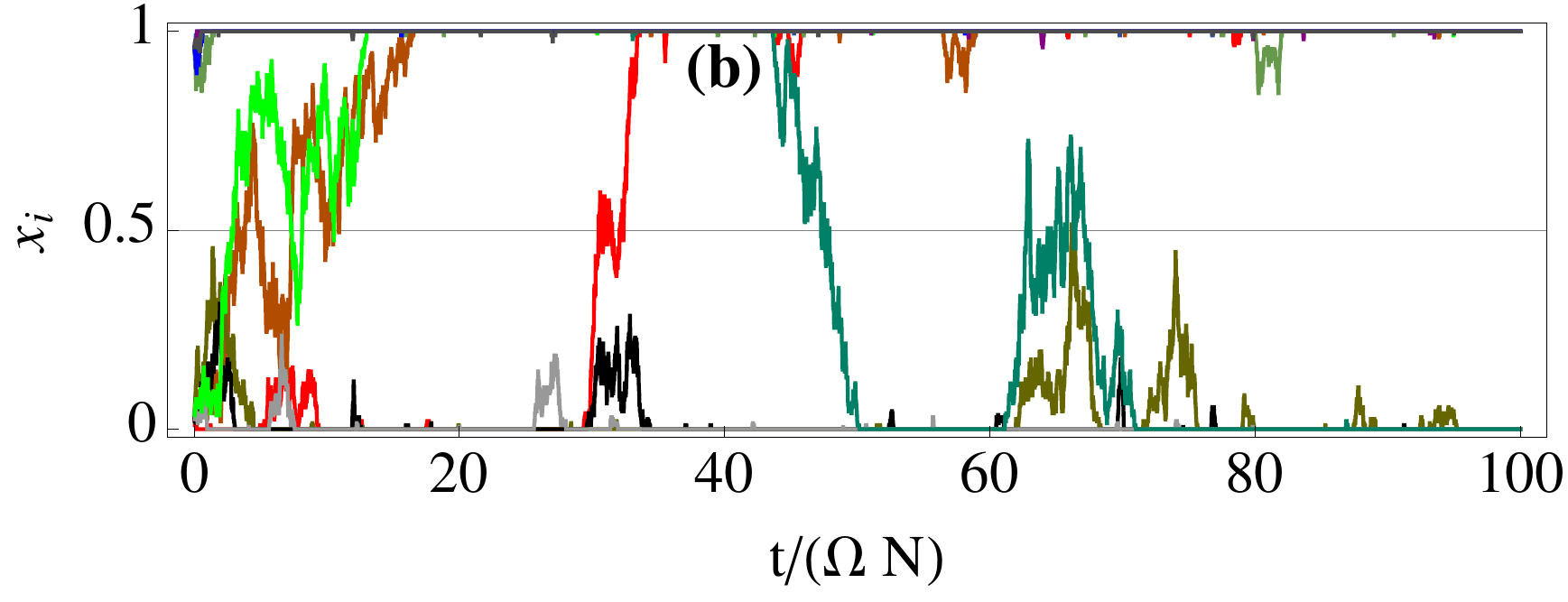}
 \caption{(Color online) Time evolution of the frequency $x_i$ of allele $A$ in the various demes (represented by different colors) of a fully-connected metapopulation consisting of $N=12$ demes with $\Omega=100$ individuals each  with (a) $m'=0.1$ or (b) $m'=0.01$ migration rate. 
 These curves are obtained from the numerical simulation of the Wright-Fisher model with balancing selection characterized by $x_\ast=0.5$ and (a) $s'=5$, (b) $s'=1$.
At time $t=0$, half of the demes have $x_i=0.05$, while the remaining ones $x_i=0.95$, which results in the ratio 
\change{$T_{\rm fix 1}/T_{\rm migr} \simeq 10^{-1}$ and $\simeq 10^{-2}$ for panel (a) and (b), respectively.}}
  \label{fig:evoluzione}
\end{figure}

\change{With a small but non-vanishing migration rate $m$, the typical time scale $T_{\rm migr} \propto 1/m'$ 
associated with the occurrence of migration can exceed the typical time  $T_{{\rm fix1}}$  needed by a single deme to reach fixation in the absence of migration (the determination of $T_{\rm fix1}$ is discussed in some detail in Appendix~\ref{app-wellmixed}) . As a result, during the time interval separating two consecutive migrations, each deme of the population rapidly evolves towards one of the ``boundary states'' $x_i=0$ or 1, which are no longer absorbing due to $m'\neq 0$, and it spends most of the time close to it. However, sometimes it happens that a different allele is received by a deme because of migration and it rapidly fixates, causing the variable $x_i$ to ``jump'' to the other boundary state.
This is illustrated in Fig.~\ref{fig:evoluzione}  (see also Fig.~1 in Ref.~\cite{prl}) which shows the time evolution of the allele frequencies $x_i(t)$ in the various demes of a population with $x_i(t=0)$ either equal to 0.05 or 0.95, for two values of migration rate (a) $m'=0.1$ and (b) $m'=0.01$ and some values of selection strength $s'$. According to Eq.~\reff{Tfix1} of Appendix~\ref{app-mft} one has $T_{\rm fix 1}/T_{\rm migr} \simeq 10^{-1}$ and $\simeq 10^{-2}$ for panel (a) and (b), respectively and, in fact, the demes in panel (a) attempt a jump between the two boundary states  more frequently than the demes in panel (b), which spend most of their time close to these boundaries. 
This dynamics can be effectively described in terms of an effective voter model, in which each deme of the metapopulation is mapped onto a voter with one of the two possible opinions which corresponds to the states $x_i=0$ or 1.}
%
%
%
%
%
%
%
Migration then acts as an effective interaction among the voters, which can influence and change each other's state.
More precisely,
with a rate  \cite{notavoter2}
\be\label{rate vot}
r=m'N/2
\ee
two randomly selected voters interact and, if they are in different states, their interaction can cause one voter (or both) to change its state.
In particular, as a consequence of the interaction between voters $i$ and $j$ with $x_i=0$ and $x_j=1$, they change state with probability $p=p(1|1/\Omega)$ and $q=p(0|1-1/\Omega)$ respectively, where $p(x'|x)$ is the probability for a single \emph{isolated} deme to reach the value $x'$ starting from an initial value $x$ before fixation occurs, and is reported in Eq.~\reff{pfix} of Appendix~\ref{app-mft}. Accordingly $p$ quantifies the probability that an isolated deme originally in the absorbing state $x_i=0$ (all individuals carry allele $B$) fixates to the opposite boundary $x_i=1$ when, because of migration, it receives an individual carrying allele $A$, such that the ensuing, single-deme fast dynamics of $x_i$ starts from the initial value $1/\Omega$. An analogous interpretation holds for $q$.
The probability that this interaction increases (respectively decreases) by one unit the number of individuals in state 1 is therefore $p(1-q)$ (respectively $q(1-p)$).
Accordingly, the rates $W_{+/-}$ at which the number of voters in state 1 increases ($+$) or decreases  ($-$) by one unit are, respectively,
\be\label{voter_rates}
 \begin{split}
  W_+&=m'Np(1-q)\bar x(1-\bar x),\\
  W_-&=m'Nq(1-p)\bar x(1-\bar x),\\
 \end{split}
\ee
where the factors $2\bar x(1-\bar x)$ account for the probability that the interacting voters are in different states.
When the number $N$ of demes is large, the master equation associated with the rates in Eq.~\reff{voter_rates} can be approximated by a Langevin equation
\be\label{Langevin_vot}
 \dot{\bar x}=\sigma_{\rm e}^{\rm vot}\bar x(1-\bar x)+\sqrt{\frac{\bar x(1-\bar x)}{N_{\rm e}^{\rm vot}}}\,\eta,
\ee
where $\sigma_{\rm e}^{\rm vot}=m'(p-q)$ is an effective directional selection coefficient, $N_{\rm e}^{\rm vot}=N/[m'(p+q-2pq)]$ is an effective population size, and $\eta$ the normalized Gaussian white noise such as the one in Eq.~\reff{eq:LG-DA}.
For small $s',m'$ and large $\Omega$, these effective coefficients reduce to
\be\label{sigmaevot}
 \begin{split}
  \sigma_{\rm e}^{\rm vot} &= 2m's(x_*-1/2),\\
  N_{\rm e}^{\rm vot} &= \frac{N\Omega}{2m'},\\
 \end{split}
\ee
and they coincide  with the coefficients $\sigma_{\rm e}$, $N_{\rm e}$ evaluated in Eq.~\reff{sigmae} and \reff{ne}, respectively. Since the expressions in Eqs.~\reff{sigmaevot}, \reff{sigmae} and \reff{ne} have been obtained 
\change{on the basis of the diffusion approximation of the dynamics of two microscopically different models, 
(i.e., the original microscopic dynamics of the island model and the effective voter model, respectively)
their agreement demonstrates that both of them correctly capture the dynamics of the system at a coarser scale. }
%
%
As it was the case in Sec.~\ref{sec:adiabatic}, the system behaves effectively as a well-mixed population with rescaled effective coefficients.
Note, however, that the deterministic term in Eq.~\reff{Langevin_vot} has the same functional form (typical of directional selection) as $M_{\rm dir}^{(0)}(\bar x)$, while the analogous of $M_{\rm symm}^{(0)}(\bar x)$ ---  the footprint of balancing selection --- is missing completely.
This is due to the fact that the specific form of the selection does not enter into the definition of the effective voter model; on the one hand, this model provides a viable approximation for the dynamics of any metapopulation with small enough migration rate but, on the other, it fails to capture some qualitative features of balancing selection.

\begin{figure}[t] 
\includegraphics[width=1.1\columnwidth]{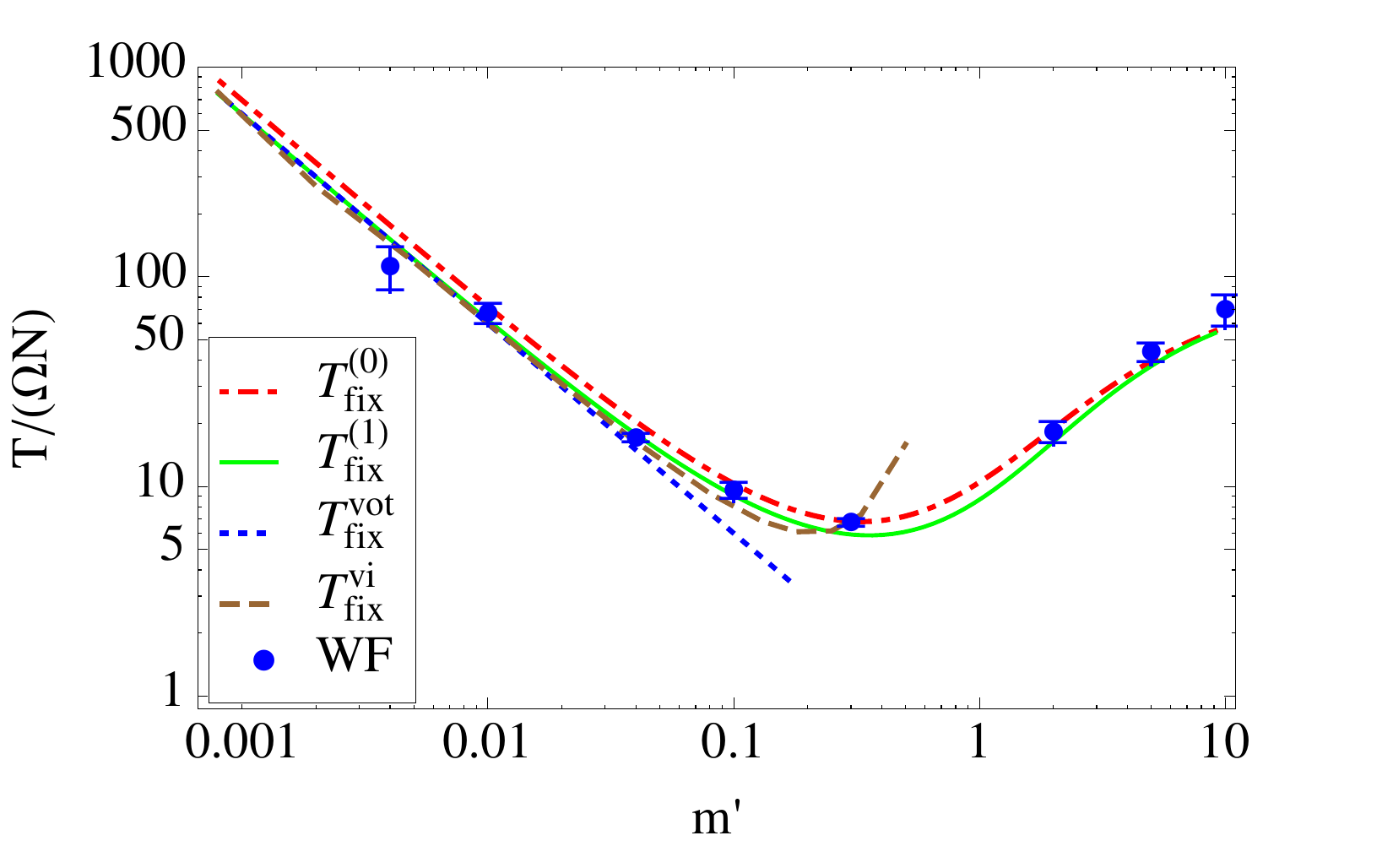}
\caption{(Color online) Mean fixation time as a function of the migration rate $m'$, for a metapopulation consisting of $N=30$ demes of $\Omega=100$ individuals each, with $s'=1$ and $x_*=1/2$. Symbols correspond to the MFT of the Wright-Fisher (WF) model obtained via numerical simulations. The red dash-dotted  and green solid curves correspond to the analytic prediction $T_{\rm fix}^{(0)}$ and its improvement $T_{\rm fix}^{(1)}$, respectively, obtained in Ref.~\cite{prl}.
The blue dashed line is the MFT $T_{\rm fix}^{\rm vot}$  [see Eq.~\reff{Tvot_symm}] of the effective voter model described in the main text. The brown dashed curve, instead, corresponds to the MFT $T_{\rm fix}^{\rm vi}$ of the voter model with an intermediate state (see Sec.~\ref{sec:voter_int}), which reproduces qualitatively the nonmonotonic behavior observed in the numerical data.
The ``lifetime'' $T_{\rm u}$ of the intermediate state introduced in Sec.~\ref{sec:voter_int} is estimated as described in Appendix \ref{app-rho}.
}
\label{fig:MFT}
\end{figure}

\change{The mean fixation time $T_{\rm fix}(m)$ of a metapopulation as a whole (see Appendix \ref{app-metapop} for its determination) depends on the initial state $x_i$ of each single deme but the mean frequency $\bar x$ actually provides an effective description of the state of the system at any time. For simplicity, in the following we focus 
on an initial state with $\bar x=1/2$: for $x_*\simeq1/2$ and a  large enough migration rate $m$, the state with  $\bar x\simeq 1/2$ actually corresponds to the metastable state onto which the population quickly relaxes from its initial state \cite{prl}.
The MFT $T_{\rm fix}^{\rm vot}$ of the voter model, instead, can be evaluated from Eq.~\reff{Langevin_vot} via the standard methods \cite{kimura-ohta} which we used in order to derive $T_{\rm fix 1}$ (see Eqs.~\reff{MFT_1deme} and \reff{Tfix1} in Appendix \ref{app-mft}) from the analogous Langevin equation~\reff{eq:LG-DA}: 
in the expression~\reff{Tfix1} for the single-deme MFT in the symmetric case $x_*=1/2$, the 
parameters $s$, $\Omega$} have to be replaced by the migration-dependent renormalized parameters $\sigma_{\rm e}^{\rm vot}$, $N_{\rm e}^{\rm vot}$, while the functions $S(a,b)$ and $F(a,b)$ in the analogous expression \reff{MFT_1deme} for generic $x_*$,
have to be replaced by the corresponding ones for the directional selection
\be
 \begin{split}
  S_{\rm vot}(a,b)&=\frac{\exp[-2\sigma_{\rm e}^{\rm vot}N_{\rm e}^{\rm vot}a]-\exp[-2\sigma_{\rm e}^{\rm vot}N_{\rm e}^{\rm vot} b]}{2\sigma_{\rm e}^{\rm vot}N_{\rm e}^{\rm vot}},\\
  F_{\rm vot}(a,b) &=\int_a^b\rmd z\int_z^1\rmd y\frac{\exp[2\sigma_{\rm e}^{\rm vot}N_{\rm e}^{\rm vot}(y-z)}{y(1-y)}.
 \end{split}
\ee
In the symmetric case $x_*=1/2$ one eventually finds
\be\label{Tvot_symm}
 T_{\rm fix}^{\rm vot}=\frac{N \log 2 }{m'p(1-p)}.
\ee
This MFT $T_{\rm fix}^{\rm vot}$ is reported in Fig.~\ref{fig:MFT} (blue dashed line) as a function of $m'$, for $\Omega=100$, $N=30$, and $s'=1$.
For small values of 
$m' \lesssim 0.03$, $T_{\rm fix}^{\rm vot}$ is in excellent agreement with the data from numerical simulations of the Wright-Fisher microscopic model (symbols, WF), with the first-order estimate $T_{\rm fix}^{(1)}$ described in Ref.~\cite{prl} (green solid line), and with the MFT $T_{\rm fix}^{\rm vi}$ (brown dashed line) obtained by introducing an intermediate state in the voter model, which we discuss further below.
%
%
%
%

\subsection{Effective voter model with an intermediate state}
\label{sec:voter_int}

In the previous section, the fixation process of the single demes was considered  to occur instantaneously and therefore each deme was supposed to be always in one of the two boundary states $x_i=0$ or 1.
However, the transition from one boundary state to the other --- triggered by the exchange of individuals between demes --- takes some time and this fact can be accounted for by introducing in the model an intermediate uncertain voter with no definite opinion.
This intermediate state  is associated with a single-deme frequency $x_i=x_{\rm u}$, where $x_{\rm u}\simeq x_*$ is an effective parameter, which depends on the optimal frequency $x_*$ and on the rates $s'$ and $m'$.
This state is supposed to be metastable, with a ``lifetime'' $T_{\rm u}$ proportional to the single-deme fixation time $T_{\rm fix1}$ reported in Eq.~\reff{MFT_1deme}; this means that the intermediate state decays with a rate $1/T_{\rm u}$ into one with definite opinion $x_i=0$ or 1.
In Appendix~\ref{app-rho} we discuss a possible heuristic estimate of the effective parameter $T_{\rm u}$.
Following the line of argument outlined in the previous subsection, and the notation introduced there, the state $x_i=1$ is reached from the intermediate state with probability $\tilde p=p(1|x_{\rm u})$.
The presence of an intermediate state is known to change completely the nature of the ordering process of the voter model \cite{voterint}. Here such a state  is introduced in order to mimic the effect of balancing selection and, as we discuss further below, it is sufficient to cause the emergence of an internal attractive point in the dynamics of $\bar x$ and nonmonotonic dependences of the MFT on the relevant parameters.

With a rate $r=m'N/2$ (the same as for the effective voter model, described in the previous subsection) 
an interaction takes place between any pair of voters. After an interaction with a voter having a different definite opinion or an indefinite one, a voter with a definite opinion can lose its own, entering the intermediate state.
%
More specifically, consider a voter $i$ in the state $x_i=0$: its interaction with a voter $j$ in a different state ($x_j=1$) consists of  the exchange of one individual between them, which introduces in the $i$-th deme an individual with allele $A$ into a background population of individuals with allele $B$ (and viceversa in the deme $j$).
The probability that the $i$-th deme, with a frequency $x_i=1/\Omega$ after the exchange, reaches the value $x_i=x_{\rm u}$ is given by $P=p(x_{\rm u}|1/\Omega)$, which represents the probability that the $i$-th voter, initially in the state $x_i=0$, reaches the intermediate state after the interaction with the $j$-th deme. Similarly, the probability that the voter $j$, initially in the state $x_j=1$, reaches the intermediate one due to its interaction with the voter $i$ is $Q=p(x_{\rm u}|1-1/\Omega)$.
Let us consider now the case of a voter in the state $x_i=0$ interacting with a voter $j$ in the intermediate one: due to this interaction, $i$ reaches the value $x_i=x_{\rm u}$ with probability $x_{\rm u}P$. Indeed, deme $i$ receives from deme $j$ an individual with allele $A$ with probability $x_{\rm u}$, in which case  the frequency $x_i$ of allele $A$ in deme $i$ reaches the value $x_u$ with probability $P$.
It is important to note that we assumed that such an interaction has no effect on the voter in the intermediate state because, for large $\Omega$, the state $x_{\rm u}\pm1/\Omega$ has almost the same fixation probability as $x_{\rm u}$ (i.e., $p(0|x_{\rm u}\pm 1/\Omega)\simeq p(0|x_{\rm u})$). 
For later purposes we emphasize here that generically $P$ increases monotonically upon increasing the selection strength $s'$, at least for $0.25 \lesssim x_*\lesssim 0.75$. This feature turns out to be crucial for understanding the nonmonotonic behavior of the MFT as a function of $s'$ (for fixed $m'$), which is discussed in detail further below.

In order to describe the dynamics of this voter model, 
we denote by $N_0$, $N_1$, and $N_{\rm u}$ the numbers of voters in states 0, 1, and $x_{\rm u}$, respectively. Since $N_0+N_1+N_{\rm u}=N$, the state of the metapopulation is fully determined by $N_0$ and $N_1$.
The rates of the possible transitions 
previously described are, in the $(N_0,N_1)$ space,
\begin{itemize}
 \item $(N_0,N_1)\overset{W_A\ }{\longrightarrow}(N_0-1,N_1),$
 \item $(N_0,N_1)\overset{W_B\ }{\longrightarrow}(N_0,N_1-1),$
 \item $(N_0,N_1)\overset{W_C\ }{\longrightarrow}(N_0-1,N_1-1),$
 \item $(N_0,N_1)\overset{W_D\ }{\longrightarrow}(N_0+1,N_1),$
 \item $(N_0,N_1)\overset{W_E\ }{\longrightarrow}(N_0,N_1+1),$
\end{itemize}
where
\be\label{rates}
 \begin{split}
  W_{A}&=\frac{m'PN_0}{N}[N_1(1-Q)+N_{\rm u}x_{\rm u}],\\
  W_{B}&=\frac{m'QN_1}{N}[N_0(1-P)+N_{\rm u}(1-x_{\rm u})],\\
  W_{C}&=\frac{m'PQN_0N_1}{N},\\
  W_{D}&=\frac{1-\tilde p}{T_{\rm u}}N_{\rm u},\\
  W_{E}&=\frac{\tilde p}{T_{\rm u}}N_{\rm u}.\\
 \end{split}
\ee
These rates define the transition matrix $W_{\vec N\to\vec N'}$ of the effective voter model with intermediate states, the stochastic evolution of which is described by the master equation \cite{Gardiner}
\be 
 \partial_t P(\vec N,t)=\sum_{\vec N'}\left[P(\vec N',t)W_{\vec N'\to\vec N}-P(\vec N,t)W_{\vec N\to\vec N'}  \right],
 \label{me-intvoter}
\ee
where $P(\vec N,t)$ is the probability to find the system in the state $\vec N=(N_0,N_1)$ at time $t$.

\subsubsection{Numerical evaluation of the MFT}

On the basis of the (forward) master  equation~\reff{me-intvoter}, a backward master equation for the fixation probability $u(\vec N,t)=P((N,0);t|\vec N;0)+P((0,N);t|\vec N;0)$ immediately follows \cite{Gardiner}
\be\label{u}
 \partial_t u(\vec N;t)=\sum_{\vec N'}\left[W_{\vec N\to\vec N'}u(\vec N',t)-W_{\vec N\to\vec N'}u(\vec N,t)  \right].
\ee
This equation can be solved numerically by introducing a time discretization $t_n=n\delta t$ (where $\delta t$ is a time interval chosen to be small enough to ensure that $W_{\vec N\to\vec N'}\delta t\ll 1$ for every pair $(\vec N,\vec N')$) and by using the finite difference approximation of the time derivative (Euler's method).
The state is described by an $(N+1)\times (N+1)$ array $u_n(N_0,N_1)$ with $N_0,N_1=0,...N$, whose entries are constrained to vanish for $N_0+N_1> N$. At each time step, the entries of $u_n$ evolve according to the discrete version of Eq.~\reff{u}.

If the system starts from a state different from the absorbing boundaries $X_1 = (N_0=0,N_1=N)$ and $X_0 = (N_0=N,N_1=0)$, the initial condition for the fixation probability is $u_0(N_0,N_1)=\delta_{N_0,N}\delta_{N_1,0}+\delta_{N_0,0}\delta_{N_1,N}$, where $\delta_{i,j} = 1$ for $i=j$, 0 otherwise. 
Since we are interested in the determination of the MFT for a system which starts from the state $(N_0=N/2,N_1=N/2)$ \cite{nota-stato-iniziale},
we focus on the quantity $U_n\equiv u((N/2,N/2), t_n)$.
The probability density $p^{\rm fix}(t)$ for reaching one of the two absorbing states as a function of the time $t$ is therefore given by the discrete derivative of $U$ for sufficiently small $\delta t$, which reads as
\be
 p_n^{\rm fix}=\frac{U_n-U_{n-1}}{t_n-t_{n-1}}.
\ee
In terms of this density,  the MFT $T_{\rm fix}^{\rm vi}$  of the voter model with intermediate state is given, for $\delta t \to 0$, by $T_{\rm fix}^{\rm vi}=\sum_{n=1}^{\infty} t_np_n^{\rm fix}$, which can be estimated as
\be\label{MFT_numerical}
 T_{\rm fix}^{\rm vi}\simeq\sum_{n=1}^{n_{\rm max}} t_np_n^{\rm fix}+T_{\rm tail},
\ee
where the term $T_{\rm tail}$ is associated with the tail of the distribution $p^{\rm fix}(t)$  for  $t > t_{\rm max}=t_{n_{\rm max}}$ and it can be  conveniently estimated by fitting $u(\vec{N},t)$ with an exponential function in the corresponding range.
In fact,  $U_n\simeq 1-e^{-\mu t_n}$ for large $t_n$, from which we obtain
\be
 T_{\rm tail}\simeq \left(t_{\rm max}+\frac{1}{\mu}\right)e^{-\mu t_{\rm max}},
\ee
where the value of $\mu$ is determined from the fit. 

Figure~\ref{fig:MFT} compares the various estimates of the MFT, as obtained from the simple voter model ($T_{\rm fix}^{\rm vot}$, blue dotted line), the voter model with intermediate states ($T_{\rm fix}^{\rm vi}$, brown dashed line) or from the lowest-order ($T_{\rm fix}^{(0)}$, red dash-dotted line) and first-order ($T_{\rm fix}^{(1)}$, green solid line) expansion in the small $s_{\rm e}/m$ parameter obtained in Ref.~\cite{prl} (with the approximation described in Section~\ref{sec:adiabatic}); symbols with errorbars, instead, correspond to the numerical results of simulations based on the Wright-Fisher model.
For small $m$, the estimates $T_{\rm fix}^{\rm vot}$ and $T_{\rm fix}^{\rm vi}$ agree with the results of simulations and with the first-order $T_{\rm fix}^{(1)}$ in the small-$s$ expansion.
It can be noticed that the introduction of the intermediate state extends to larger values of $m$ the range within which the approximation is accurate and, more importantly, it makes the model able to capture qualitatively the nonmonotonic behavior of the MFT as a function of $m$.
This demonstrates that the existence of the intermediate (metastable) state plays a crucial role in determining the emergence of the nonmonotonicity in the mean fixation time, as it was argued in Ref.~\cite{prl}.

\begin{figure}[t] 
\includegraphics[width=\columnwidth]{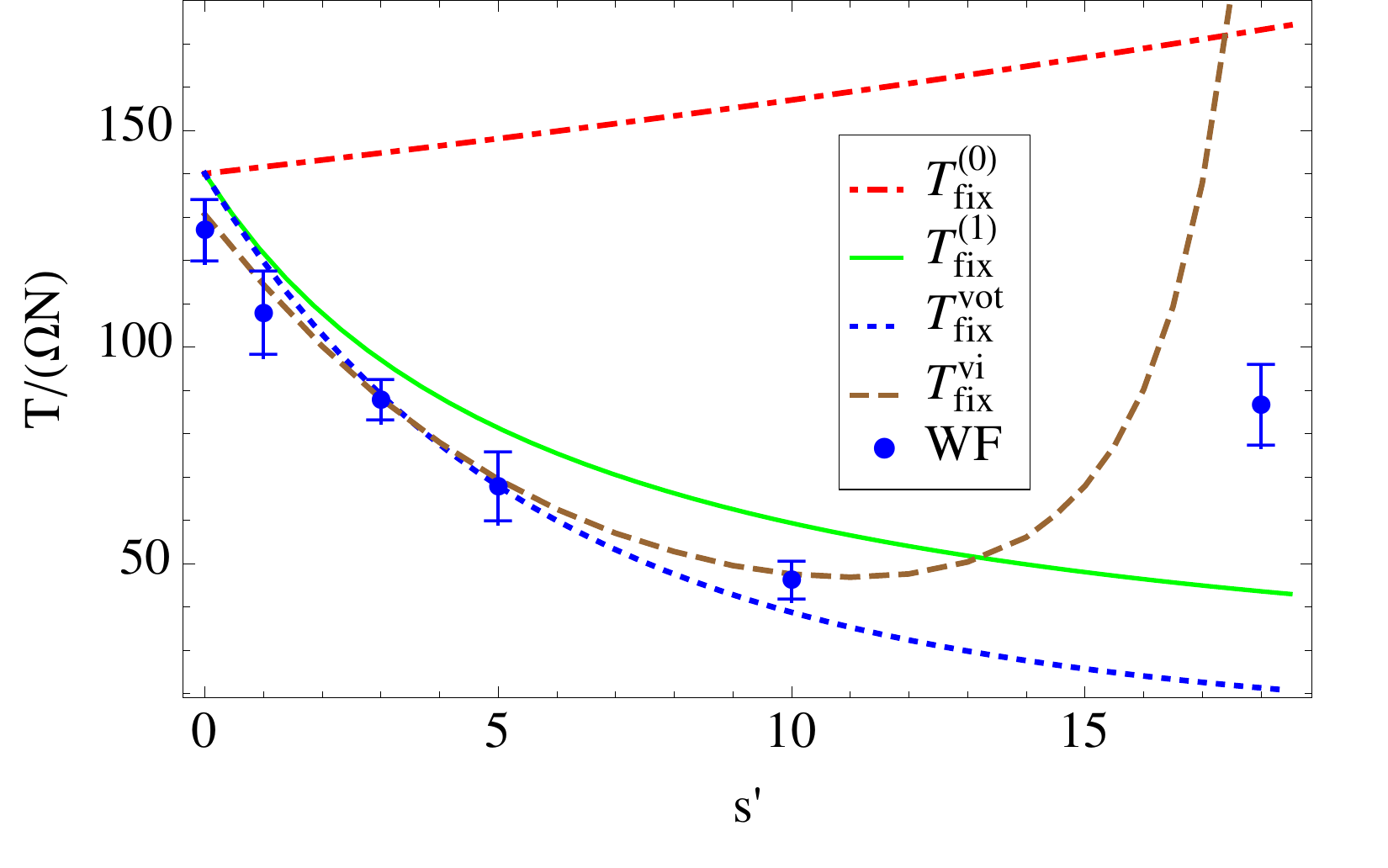}
\caption{(Color online) MFT for a metapopulation of $N=30$ demes with $\Omega=100$ individuals, $m'=0.005$, and $x_*=1/2$. 
Symbols correspond to the MFT of the Wright-Fisher (WF) model obtained via numerical simulations. The red dash-dotted  and green solid lines correspond to the analytical predictions $T_{\rm fix}^{(0)}$ and $T_{\rm fix}^{(1)}$ obtained in Ref.~\cite{prl}.
 The blue dashed line is the MFT $T_{\rm fix}^{\rm vot}$ 
[see Eq.~\reff{Tvot_symm}] of the effective voter model described in the main text. The brown dashed curve, instead, corresponds to the MFT $T_{\rm fix}^{\rm vi}$ of the voter model with an intermediate state (see Sec.~\ref{sec:voter_int}), which reproduces qualitatively the nonmonotonic behavior observed in the numerical data.
We have estimated the lifetime $T_{\rm u}$ of the intermediate state as described in Appendix \ref{app-rho}.
The scenario presented here carries over to different choices of $m'\lesssim0.05$.
}
\label{fig:MFT2}
\end{figure}

In Fig.~\ref{fig:MFT2} we report the MFT as a function of the selection rate $s'$ for a fixed small value of the migration rate $m'=0.005$. It can be noticed that  $T_{\rm fix}^{\rm vi}$ from Eq.~\reff{MFT_numerical} is in excellent agreement with the results of the numerical simulations of the Wright-Fisher model (symbols) also for quite large values of the selection rate $s'$; the introduction of the intermediate state in the voter model significantly improves the accuracy of the approximation compared to both $T_{\rm fix}^{(0)}$ and $T_{\rm fix}^{(1)}$ discussed in Ref.~\cite{prl}
and to $T_{\rm fix}^{\rm vot}$ of the voter model without intermediate state.

Since balancing selection tends to push all the demes towards the configuration with allele frequency $x_*$ which is far from the boundaries (at least for $x_*\simeq 1/2$), it is heuristically expected to cause a slowing down of fixation and therefore to increase the MFT; however, Fig.~\ref{fig:MFT2} shows that this is not always the case and in fact the MFT plotted there displays a nonmonotonic behavior as a function of the selection rate.
This nonmonotonicity appears for  small enough $m'$ and it can be rationalized on the basis of the 
\change{effective voter model} 
with intermediate states. In fact, the MFT $T_{\rm fix}^{\rm vi}$ is expected to be proportional to the mean time $T_{\rm change}$ that a voter needs to change its opinion, which can be estimated as $T_{\rm change} \simeq T_{\rm int}+T_{\rm u}$, where $T_{\rm int}$ is the time scale associated with an interaction able to drive a voter initially in states 0 or 1 into the intermediate one $x_{\rm u}$ with ``lifetime'' $T_{\rm u}$.
Since a voter interacts with a typical rate $m'$ and, after this interaction, it reaches the intermediate state $x_{\rm u}$ with probability $P$, the rate $T_{\rm int}^{-1}$ associated with the transitions towards the intermediate state is given by $T^{-1}_{\rm int}\simeq m'P$, so that
\be
 T_{\rm change}\simeq \frac{1}{m'P}+T_{\rm u}.
\ee
For small $s'$, the mean time $T_{\rm change}$ is predominantly determined by the term $1/(m'P)$, which in fact increases upon decreasing $s'$, while in the opposite limit of large $s'$ it is actually determined by $T_{\rm u}$, which increases upon increasing $s'$.
The interplay between these two terms results in the nonmonotonic dependence of $T_{\rm change}$ --- and therefore of $T_{\rm fix}^{\rm vi}$ --- on $s'$.
However, upon further increasing $s'$, it is no longer correct to assume that each deme spends a large part of its time into a boundary state, and therefore in this regime one cannot expect $T_{\rm fix}^{\rm vot}$ and $T_{\rm fix}^{\rm vi}$ to reproduce accurately the corresponding results of numerical simulations of the Wright-Fisher model; nonetheless $T_{\rm fix}^{\rm vi}$ still captures the qualitative behavior of  $T_{\rm fix}$ of such a model, as it is clearly seen in Fig.~\ref{fig:MFT2} by comparing the symbols (numerical simulations) with the dashed line.

%
%
%
%

\subsubsection{Effective equation for the mean frequency $\bar x$}
\label{sec:effective_eq}

When the migration rate $m$ is small (and therefore the interaction rate $r$ among the voters is small --- see Eq.~\reff{rate vot})
only a relatively small fraction of voters is in the intermediate state, i.e., $N_{\rm u}\ll N$.
For large $N$, the evolution of $\bar x$ is expected to be slow compared to that of $N_{\rm u}$, because every interaction causes a change 
$\Delta N_{\rm u}=\pm 1$ of $N_{\rm u}$, but only a change $\Delta \bar x \lesssim 1/N$ of $\bar x \sim 1$, so that the relative variation $|\Delta N_{\rm u}|/N_{\rm u}$ of the former is significantly larger than that of the latter $|\Delta\bar x|/\bar x\ll |\Delta N_{\rm u}|/N_{\rm u}$.
This time scale separation allows us to consider $N_{\rm u}$ as a fast fluctuating variable on the time scale which characterizes the dynamics of  
$\bar x$. Conversely, $\bar x$ can be considered as a slowly varying (or almost constant) parameter on the time scale of the dynamics of 
$N_{\rm u}$.

For the sake of simplicity, we focus here on the case of symmetric balancing selection ($x_*=1/2$ and therefore $x_{\rm u}=1/2$, $\tilde p=1/2$, and $P=Q$), but the discussion below can be straightforwardly generalized to the non-symmetric case, with similar conclusions. 
Considering the characteristic time scale over which the number $N_{\rm u}$ of voters in the intermediate state evolves, 
an estimate of the 
mean 
$\langle N_{\rm u}\rangle$ in the large-$N$ and small-$m$ limit can be found by solving the stationary master equation which describes the evolution of $N_{\rm u}$ (see Appendix \ref{app-Nu} for details)
\be\label{N_umedio}
 \langle N_{\rm u}\rangle \simeq\frac{2m'NPT_{\rm u}\bar x(1-\bar x)}{1-m'PT_{\rm u}/2}.
 \ee
Note that, as expected, the mean number $\langle N_{\rm u}\rangle$ of voters in the intermediate state vanishes as $m'\to 0$.
On the time scales over  which $\bar x$ varies, we can approximate $N_{\rm u}$ with its mean $\langle N_{\rm u}\rangle$ [see Eq.~\reff{N_umedio}]; this allows us to write an effective Langevin equation for the evolution of the mean frequency $\bar x$, which reads as
\be\label{Langevin_vi}
 \dot{\bar x}=s_{\rm vi}\bar x(1-\bar x)(1/2-\bar x)+\sqrt{\frac{\bar x(1-\bar x)}{N_{\rm vi}}}\,\eta(t),
\ee
 for large $N$ and 
$\Omega$ and small $m$ (see Appendix \ref{app-xbar} for a detailed derivation),
where  $s_{\rm vi} =m'^2P^2T_{\rm u}/(1-m'PT_{\rm u}/2)$ is an effective selection coefficient and $N_{\rm vi} =N/(m'P)$ is an effective population size.
These effective parameters coincide with $s_{\rm e}$ and $N_{\rm e}$ (see Eqs.~\reff{se} and \reff{ne}), respectively, for large $\Omega$ and small $s'$.
\change{Since $s_{\rm e}$, $N_{\rm e}$ and $s_{\rm vi}$, $N_{\rm vi}$ have been obtained 
by considering the diffusion approximation of the dynamics of two microscopically different models (the original island model and the effective voter model, respectively),
their agreement show the emergence of a coherent effective dynamics at a coarser scale.}
Note that the deterministic term in Eq.~\reff{Langevin_vi} has an internal attractive point $\bar x_* = 1/2$, which is the footprint of balancing selection,  and this means that the intermediate state $x_{\rm u}$ is the crucial ingredient in order to capture the main  features of balancing selection; note also that $s_{\rm vi} \propto m'^2$ for small $m'$, where indeed the approximation in Eq.~\reff{Langevin_vot} is accurate.

\section{Summary and conclusions}
\label{sec:conclusions}

Balancing selection is a major mechanism responsible for promoting and maintaining  biodiversity, as it favors the coexistence of different alleles in the same population. Under balancing selection, the evolution of a population is characterized by the emergence of a long-lived metastable state (at least for sufficiently strong selection), which is eventually destabilized by stochasticity (genetic drift). 
When the population is subdivided in many subpopulations of equal size and features, migration interacts with balancing selection and with genetic drift to determine the ultimate fate of the population. In Ref.~\cite{prl} we noticed that the interplay of these evolutionary ``forces'' leads to the emergence of a separation of time scales between the global and the local dynamics, which can be used in order to develop an approximate description of the dynamics and therefore to determine the fixation properties of the population as a whole.
In the present work, we have extended this approach in two directions. First, we have shown that, contrary to the heuristic expectation, balancing selection actually speeds up fixation with respect to a neutral model (i.e., a model without selection) if the allele frequency $x_*$ promoted by balancing selection in the coexistence state is close to extinction of one of the alleles. 
This phenomenon, already observed in Ref.~\cite{robertson} for well-mixed populations, carries over to a subdivided population, where, in addition, it is responsible for the emergence of a phase transition in the limit of an infinite number $N$ of subpopulations, each of finite size $\Omega$.
We heuristically explain this behavior in Sec.~\ref{sec:PT} by decomposing the effect of asymmetric balancing selection on the evolution of the mean allele frequency $\bar x$ as a sum of a symmetric balancing selection term $M_{\rm symm}(\bar x)$ and a directional term $M_{\rm dir}(\bar x)$, which favor coexistence and  fixation, respectively. In fact, it turns out that $M_{\rm dir}(\bar x)$  becomes stronger than $M_{\rm symm}(\bar x)$ as $x_*$ approaches one of the two boundaries $x_*=0$ or 1, corresponding to the extinction of one of the alleles. 
\change{The results presented here are compatible 
with the critical behavior expected for the DP2 universality class within the mean-field approximation \cite{hinrichsen}.}
It is then possible to characterize in detail the mean fixation time of finite populations as a function of the migration rate $m$ and of the other relevant parameters, the selection strength $s'$ and the optimal frequency  $x_*$. While the perturbative results provided in Ref.~\cite{prl} are limited either to fast migration or to moderate balancing selection, we have shown here how to extend them to slow migration and larger values of selection strength. In fact, a metapopulation with a small migration rate $m$ can be effectively described at a coarser level by a voter model with an interaction rate proportional to the migration rate $m$ in the metapopulation. In Sec.~\ref{sec:voter} we demonstrate that the MFT of this effective voter model correctly reproduces the one of the original metapopulation  for slow migration. 
However, the standard voter model fails to reproduce some qualitative features of the MFT of the subdivided population, which are recovered once we introduce into the model an additional intermediate state, corresponding to a voter with no definite opinion. This intermediate state turns out to be crucial for reproducing the nonmonotonic behavior of the MFT as a function of the migration rate, originally observed in Ref.~\cite{prl} (see Fig.~\ref{fig:MFT} here); in addition, this model provides prediction for the MFT in good quantitative agreement with simulations up to larger values of the migration rate.
We have also shown that an analogous nonmonotonic behavior of the MFT emerges as a function of the selection strength for a sufficiently small and fixed migration rate, in qualitative and partially quantitative agreement with numerical simulation of the microscopic model (see Fig.~\ref{fig:MFT2}). 

In summary, the three-state effective model proposed here provides a coarser description of the collective behavior of the metapopulation that is useful in order to understand the mechanisms underlying  the emerging phenomena observed in the population. Such a description is expected to carry over to other population models in which it is possible to identify a separation of time scales between local and global dynamics. In this respect, the metapopulation considered here has a very simple internal structure (it is a fully-connected graph), therefore it would be important to investigate whether the features discussed above are present on more general networked (or even spatially-embedded) systems and to understand their potential interplay with other dynamical phenomena, such as diffusion and coarsening.
Since the validity of the \change{effective voter models} presented in this work only relies on the ``slowness'' of migration compared to the other forces driving the dynamics, these approximations could  be adapted to various networks and spatial lattices, or even to different form of the inter-deme dynamics, as long as the migration between the demes is slow enough.


\appendix

\section{Diffusion approximation for the microscopic models}\label{app-diffusion}

\change{Given a microscopic model (e.g., the Wright-Fisher model) with certain transition rates, the evolution equation for the probability $P(x,t)$ of finding the population in a certain configuration $x$ at time $t$ can be written in the form of a master equation. The \emph{diffusion approximation} then consists in approximating this equation with a Fokker-Planck (or equivalently, a Langevin) equation, i.e., with a diffusion equation with suitable $x$-dependent drift and diffusion coefficients $\mu$ and $v$, respectively:
\be\label{FP_Moran1}
   \partial_t P(x,t)=- \partial_{x}[\mu(x)P(x,t)] +\frac{1}{2}\partial_{x}^2[v(x)P(x,t)].
\ee
In particular, only the mean value $\langle \Delta x \rangle \propto  \mu(x)$ and the variance $\langle (\Delta x)^2 \rangle  \propto v(x)$ of the change $\Delta x$ per unit time of the variable $x$ resulting from the implementation of these microscopic dynamical transitions are accounted for in the evolution equation.} 
Although this approach is known to fail in some cases, for example for the susceptiple-infected-susceptible (SIS) model of epidemiology (see, e.g., Ref.~\cite{sis}), it turns out to be quite accurate for the Wright-Fisher and Moran models discussed in the present work \cite{prl,diffusion}.
\change{Below we illustrate how to implement the diffusion approximation for the Wright-Fisher and Moran models; in addition, we also discuss how one can modify the microscopic rules of these two stochastic models in order to account for the migration occurring within the island model, which takes the form of exchanges of individuals among subpopulations.}

\subsection{Wright-Fisher model}\label{app-diffusionWF}

For a well-mixed population of sufficiently large size $\Omega$, one can readily calculate the mean and the variance of the change $\Delta x$ per unit time of the allele frequency $x$ from the binomial sampling probability $p_{\rm r}(x)$ given in Eq.~\eqref{prWF}, obtaining 
\be
\begin{split}
 \mu(x)  &=  \langle \Delta x\rangle =\tilde s x(1-x)+O(\tilde s^2),\\
 v(x)  &=  \langle (\Delta x)^2\rangle = \frac{x(1-x)+O(\tilde s)}{\Omega}+O(\tilde s^2).
\end{split}
\label{eq:mu-v-WF}
\ee

\change{A standard way to account for migration in the Wright-Fisher model is to modify the probability  $p_{\rm r}(x)$ with which a new generation is sampled.  In fact, in the migration process which occurs between two subsequent generations, a mean number $m\Omega$ of randomly chosen individuals leaves each deme and it is then 
randomly redistributed in the other demes. One can effectively think of all the individuals leaving the $N$ demes as merging in a sort of ``reservoir'', with a mean of $m\Omega N$ individuals, and a mean fraction of type-$A$ individuals determined by the inter-deme mean frequency 
$\bar x={N}^{-1}\sum_{i=1}^Nx_i$. Individuals are then randomly chosen from this reservoir in order to replace those which migrated from each deme.} 
\change{Accordingly, the mean fraction of type-$A$ individuals arriving in the $i$-th deme because of this redistribution is $m\Omega \bar{x}$, while the mean fraction of type-$A$ individuals leaving deme $i$ is $m\Omega x_i$. As a result of migration, in each deme of the metapopulation, the original fraction $x_i$ has changed into $m\bar{x} + (1-m)x_i$.
This simplified description of the migration process neglects fluctuations in the number and in the composition of migrants during each generation and 
in fact, it accounts for migration only by modifying the probability $p_r$ that a new individual carries allele $A$ \cite{prl,cherry-wakeley}.
In particular, due to migration, the probability $p_r(x_i,\bar x)$ that a new individual in deme $i$ carries allele $A$ acquires a dependence on the mean frequency $\bar x$, and it has the form 
\be\label{pWF}
p_{\rm r}(x_i,\bar{x}) = \frac{(1+\tilde s_i)[m \bar x+(1-m)x_i]}{1+\tilde s_i [m \bar x+(1-m)x_i]},
\ee
where $\tilde s_i = \tilde s(x = m\bar x+(1-m)x_i)$ is the value of the function $\tilde s(x)$ [for balancing selection $\tilde s(x)=s(x_*-x)$] 
evaluated at $x=m\bar x+(1-m)x_i$.
At this point, it is easy to extend the diffusion approximation to a subdivided population (island model) characterized by the binomial sampling probability given in Eq.~\eqref{pWF},  obtaining the mean and the variance of the change in the fraction of $A$-type individuals of a subpopulation $i$ as   
\be
\begin{split}
 \mu(x_i)  &=  \langle \Delta x_i\rangle =\tilde s x_i(1-x_i)+m(\bar x-x_i)+O(\tilde s^2,\tilde s m),\\
 v(x_i)  &=  \langle (\Delta x_i)^2\rangle = \frac{x_i(1-x_i)+O(m,\tilde s)}{\Omega}+O(\tilde s^2,m^2,m\tilde s).
\end{split}
\label{eq:mu-v-WF}
\ee
These expressions lead directly to the Langevin equation~\reff{Langevin_single-deme}.}

\subsection{Moran model}
\label{app-diffusionM}

\change{In a well-mixed population, the time evolution of the probability distribution $P(x,t)$ of the frequency $x$ can be determined from the corresponding master equation with the rates given by Eq.~\eqref{rateMoran1}. For large $\Omega$ and in the limit of continuous time $\delta t \to 0$ (where $\delta t$ denotes the duration of a step in the dynamics of the Moran model),  standard expansions, such as the Kramers-Moyal expansion \cite{Gardiner}, lead to Eq.~\reff{FP_Moran1} 
in which the drift $\mu$ and the variance $v$ are given by
\be\label{drift_var}
\begin{split}
  \mu(x) &=\frac{W_{1}-W_{-1}}{\Omega\,\delta t}=\frac{\tilde s}{2} x(1-x)+O(\tilde s^2),\\
  v(x) &=\frac{W_{1}+W_{-1}}{\Omega^2\,\delta t}=\frac{x(1-x)+O(\tilde s)}{\Omega}.
\end{split}
\ee}

\change{In a metapopulation consisting of $N$ demes, instead, the allele frequency $x_i$ of each deme $i$ can additionally change, during each step of the evolution,  because of migration.} 
More precisely, the probability that in the $i$-th deme the number $\Omega_A$ of individuals carrying allele $A$ increases (decreases) by one unit is given by $W_1^{\rm m}\delta t $ ($W_{-1}^{\rm m}\delta t $), where $\delta t=O(\tau_g/\Omega)$ is the duration of the evolutionary step 
and the rates are given by
\be
 \begin{split}
  W_{+1}^{\rm m} &=m\bar x(1-x_i),\\
  W_{-1}^{\rm m} &=m(1-\bar x)x_i,\\
 \end{split}
\ee
where we neglect the $O((m\delta t)^2)$ probability that more than one individual per deme migrates within one evolutionary step.
Because of the simultaneous action of death, reproduction, and migration, the probability $Q_k$ to have a change $k = \pm1,\pm2$ in the number of alleles $A$ in the $i$-th deme between two subsequent generations is given by 
\be
 \begin{split}
  Q_{+2} &= W_{+1}\delta t\,W_{+1}^{\rm m}\delta t,\\
  Q_{+1} &= W_{+1}\delta t\left[1-(W_{+1}^{\rm m}+W_{-1}^{\rm m})\delta t\right]\\
  &\quad +\left[1-(W_{+1}+W_{-1})\delta t\right]W_{+1}^{\rm m}\delta t,\\
  Q_{-1} &= W_{-1}\delta t\left[1-(W_{+1}^{\rm m}+W_{-1}^{\rm m})\delta t\right]\\
  &\quad +\left[1-(W_{+1}+W_{-1})\delta t\right]W_{-1}^{\rm m}\delta t,\\
  Q_{-2} &= W_{-1}\delta t\,W_{-1}^{\rm m}\delta t.\\
 \end{split}
\ee
\change{If one neglects the contribution of order $O(\delta t^2)$, and for $\Omega\gg1$, the evolution turns out to be described by the rates 
 \be
 \begin{split}
  \label{rateMoran2}
  W_{+1} &= (1+\tilde s)x_i(1-x_i)/(1+\tilde s x_i)+m\bar x(1-x_i),\\[1mm]
  W_{-1} & = x_i(1-x_i)/(1+\tilde s x_i)+m(1-\bar x)x_i,
\end{split}
\ee 
associated with the $Q_k$'s discussed above.
The  probability distribution $P(\{x_i\},t)$ describing the evolution of the state of the whole metapopulation satisfies a master equation that, in the diffusion approximation, gives the multivariate Fokker-Planck equation}
\be\label{FP_Moran}
 \begin{split}
   \partial_t P(\{x_i\},t)&=-\sum_{j=1}^N\partial_{x_j}[\mu(x_j)P(\{x_i\},t)]\\
   &+\frac{1}{2}\sum_{j=1}^N\partial_{x_j}^2[v(x_j)P(\{x_i\},t)],
 \end{split}
\ee
in which the drift $\mu$ and the variance $v$ are given by
\be\label{drift_var}
\begin{split}
  \mu(x_i) &=\frac{W_{1}-W_{-1}}{\Omega\,\delta t}=\frac{\tilde s}{2} x_i(1-x_i)+\frac{m}{2}(\bar x-x_i)+O(\tilde s^2),\\
  v(x_i) &=\frac{W_{1}+W_{-1}}{\Omega^2\,\delta t}=\frac{x_i(1-x_i)+O(\tilde s,m)}{\Omega},
\end{split}
\ee
where we have chosen the temporal step to be $\delta t = 2/\Omega$.
With this choice of time scales, the resulting genetic drift $v(x_i)$ for small  $\tilde s$ and $m$ is the same as the one of the  Wright-Fisher model for a population of the same size, see Eq.~\reff{eq:mu-v-WF}.
Note that, in order to recover the same expression also for the drift $\mu(x_i)$, it is necessary to rescale the migration and the selection coefficients as $m\to 2m$ and $\tilde s\to2\tilde s$, respectively. 
Equation~\eqref{FP_Moran} is nothing but the Fokker-Planck equation associated with the set of $N$ single-deme Langevin equations \eqref{Langevin_single-deme}, which, as we argued above, also describe the dynamics of the Wright-Fisher model in the presence of migration.

\section{\change{Mean fixation time and fixation probability}}
\label{app-mft}

\subsection{\change{Well-mixed population}}
\label{app-wellmixed}

\change{The time required in order to reach fixation in a well-mixed population of size $\Omega$ is a stochastic variable. Its mean, i.e., the MFT $T_{\rm fix1}$, can be evaluated by standard 
methods \cite{kimura-ohta} in the diffusion approximation with drift $\mu(x)$ and variance $v(x)$ defined in Section~\ref{sec:micro-meso} . Let us define the quantity $G(x) = \rme^{-\int_0^x \frac{2\mu(x')}{v(x')} dx'}$  and the functions $S(a,b)$ and $F(a,b)$ 
\be
 \begin{split}
  S(a,b) &=\int_a^b \rmd x\ G(x)   = \int_a^b\rmd x\exp\left[-s'x(2x_*-x)\right],\\
  F(a,b) &= \int_a^b\rmd z\int_z^1\rmd y \frac{G(z)}{\Omega v(y)G(y)} \\
   & =  \int_a^b\rmd z\int_z^1\rmd y\frac{\exp\left\{s'\left[y(2x_*-y)-z(2x_*-z)\right]\right\}}{y(1-y)},
 \end{split}
\ee
where we conveniently introduced the rescaled selection coefficient $s'=\Omega s$.
The mean fixation time reads as
\be\label{MFT_1deme}
 \frac{T_{\rm fix1}(x)}{\Omega}=\frac{2\left[S(x,1)F(0,x)-S(0,x)F(x,1)\right]}{S(0,1)},
\ee
where $x$ is the initial condition.
In the symmetric case $x_*=1/2$,  
Eq.~\reff{MFT_1deme} reduces to
\be\label{Tfix1}
  \frac{T_{\rm fix1}(x)}{\Omega}=
  \int_{(1-2x)^2}^1\rmd u\int_0^1 \rmd z\frac{\rme^{s'u(1-z^2)/4}}{1-uz^2}.
\ee
As expected, $T_{\rm fix1}(x)$ vanishes if the initial condition $x$ corresponds to one of the two absorbing states, $x=0$ or 1, 
while it reaches smoothly its maximum value as the initial condition moves towards $x=1/2$. In this work (as well as in Ref.~\cite{prl}) we focused on the initial condition $x=1/2$, which, for $x_*\simeq1/2$ and $s$ large enough, corresponds to a long lived metastable state promoted by balancing selection. }

\change{Note that starting from an initial value $x_0$, the frequency $x$ does not typically visit the whole interval of possible values $x\in (0,1)$ during its evolution because of the presence of absorbing states which cause fixation: in fact, the probability $p(x_1|x_0)$ that the population reaches the value $x_1$ during the evolution which precedes fixation can be evaluated via a standard procedure \cite{pfix} and it reads
\be\label{pfix}
 p(x_1|x_0)=
  \left\{
  \begin{array}{ll}
    \displaystyle\frac{\int_0^{x_0}\rmd y\, G(y)}{\int_0^{x_1}\rmd y\, G(y)}&{\rm if}\ x_1>x_0, \vspace{1mm}\\
    \displaystyle\frac{\int_{x_0}^1\rmd y\, G(y)}{\int_{x_1}^1\rmd y\, G(y)}&{\rm if}\ x_1<x_0. 
  \end{array}\right.
\ee}

\subsection{\change{Metapopulation}}
\label{app-metapop}

\change{As argued in the main text, for a metapopulation, the mean fixation time $T_{\rm fix}(m)$ effectively depends on the mean frequency $\bar x$ and for simplicity, we focus on a symmetric initial state with $\bar x=1/2$ \cite{prl}. $T_{\rm fix}(m)$ differs from the one of a single deme  $T_{\rm fix1}$ also in the absence of migration, i.e., for $m=0$, when each deme evolves independently of the others. In this case, the mean time $T_{\rm fix}(m=0)$ required by the overall population to reach one of the two absorbing states is given by the mean time necessary for all demes to reach it, after which no evolution occurs within the metapopulation. This is given by the maximum of the single-deme fixation times calculated over $N$ demes, and it turns out to be $T_{\rm fix}(m=0)\simeq T_{\rm fix 1}\log N$ \cite{prl}.
For non-zero migration, the approximation proposed in Ref.~\cite{prl} and summarized in Sec.~\ref{sec:adiabatic} makes possible to calculate the mean fixation time for the metapopulation using formula \reff{MFT_1deme} with modified expressions for the drift and the variance.  
In particular, by using the drift and the variance reported in Eqs.~\reff{m0x} and \reff{v0x}, one obtains
\be 
T_{\rm fix}^{(0)} = \frac{N_{\rme}}{2}\int_0^1 dy \int_0^1 \frac{{\rm e}^{s_{\rm e} N_{\rm e} y (1-z^2)/4}}{1-y z^2},
\ee
i.e., Eq.~(5) of Ref.~\cite{prl}. Analogously, by using the improved approximations for the drift and the variance discussed therein, one obtains the estimate $T_{\rm fix}^{(1)}$ provided in the supplemental material of that work.}

\section{Estimate of the ``lifetime'' $T_{\rm u}$ of the intermediate state}\label{app-rho}

\begin{figure}[h!]
\centering
\includegraphics[width=\columnwidth]{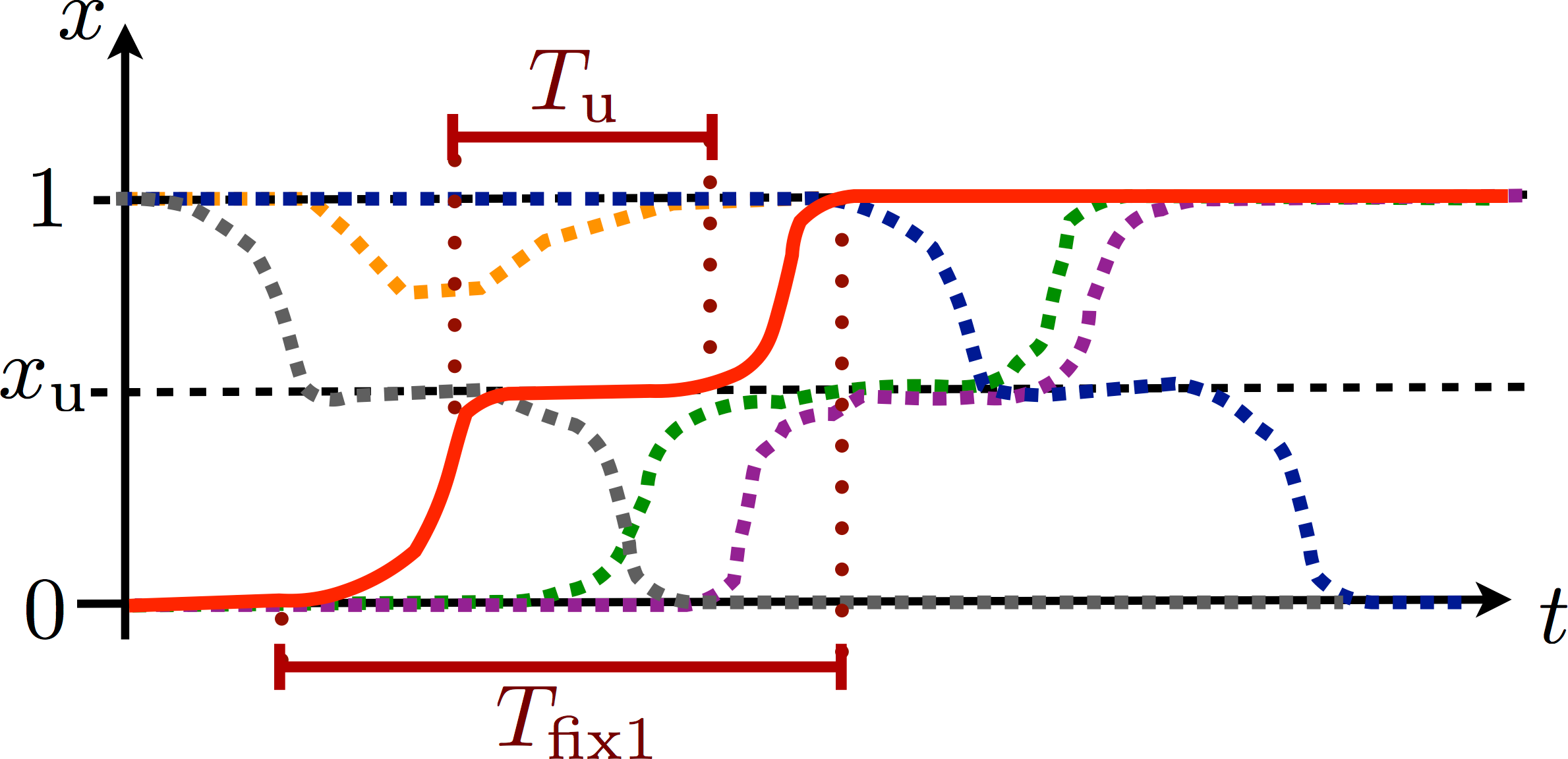}
\caption{(Color online) Schematic representation of the evolution of the frequencies $x_i$ of various demes (represented by different colors) in a metapopulation with small migration rate $m'\ll T_{\rm fix1}^{-1}$. Under this assumption, the possible transition between the two boundary states is triggered with a typical time scale $\sim 1/m'$ which is much longer than the one $T_{\rm fix1}$ taken by the transition itself, where  $T_{\rm fix 1}$ is the single-deme MFT. Part of this time, corresponding to the ``lifetime'' $T_{\rm u}$, is spent by the various demes in the vicinity of the intermediate state.}
\label{fig:disegno}
\end{figure}

%
\change{If the  migration rate $m'$ is sufficiently small, such that  $T_{\rm fix1} \ll T_{\rm migr} \propto 1/m'$,}
each deme spends most of its time into one of the two boundary states, until it receives, due to migration, one individual different from the majority. In turn, this individual triggers an attempt to leave the boundary state which leads to the intermediate one $x_{\rm u}$ (and possibly to the opposite boundary) with an overall rate $m'P$, as discussed in Sec.~\ref{sec:voter_int} \change{(see also Fig.~\ref{fig:evoluzione})}.
Under our assumption of small migration rate, this transition takes place before any other individual is exchanged by the deme  with the rest of the population, and therefore it occurs as in an isolated deme, i.e., it takes a mean time $T_{\rm fix 1}$. 
During this transition, the deme will spend a mean time $T_{\rm u}$ close to  the intermediate state $x_{\rm u}$ before reaching the final boundary.
Figure~\ref{fig:disegno} provides a schematic representation of 
the time evolution of the allele frequencies of the various demes (indicated by solid and dashed lines of different colors) in the regime described above.
 In particular, the deme represented by the solid line has received an individual with allele A from another deme and has fixed it after a time $T_{\rm fix 1}$, of which $T_{\rm u}$ spent close to the intermediate state promoted by balancing selection. 
  Other demes of the metapopulation (indicated by dashed lines) evolve similarly, i.e., they move from one boundary state to the other, with possible unsuccessful attempts.
%
%
%
%

In order to estimate $T_{\rm u}$, we focus on the ratio $\rho=T_{\rm u}/T_{\rm fix1}$, i.e., on
the fraction of the time spent outside the boundaries in which the deme is actually close to $x_{\rm u}$; in the quasi-stationary state associated with a certain value of the mean frequency $\bar x$ (and described by the density $P_{\rm qs}(x|\bar x)$),  $\rho$ can be estimated as the ratio of the corresponding probabilities, i.e., of the probability to find a deme close to the intermediate state (that is $x\simeq x_{\rm u}$) to the one of finding it outside the boundaries.
In order to specify properly the condition of $x_i$ being ``close to'' $x_{\rm u}$, hereafter we focus for simplicity on the symmetric case with $x_*=x_{\rm u}=1/2$.
In the effective voter model with the intermediate state, the continuous interval of states $x\in [0,1]$ is represented by the three coarse states $\{0,1/2,1\}$. It is then natural to associate to every value of $x$ in the interval $[0,1]$ its closest representative state.
With this definition, the probability $P_{\rm u}$ to find the deme close to the intermediate state is
\be
P_{\rm u}(\bar x)=\int_{1/4}^{3/4}{\rm d}x\,P_{\rm qs}(x|\bar x).
\ee
On the other hand, the probability to find the deme outside the boundary states
can be estimated as
\be
P_{\rm non-fix}(\bar x)=\int_{1/\Omega}^{1-1/\Omega}\rmd x\,P_{\rm qs}(x|\bar x),
\ee
where a minimal distance $1/\Omega$ of the deme frequency from a boundary corresponds to having one individual different from the background.
The ratio $\rho$ can therefore be approximated as
\be\label{rho_eq}
\rho(\bar x,s',m')=\frac{P_{\rm u}(\bar x)}{P_{\rm non-fix}(\bar x)}.
\ee
%
%
\begin{figure}[h!]
\centering
\includegraphics[width=1\columnwidth]{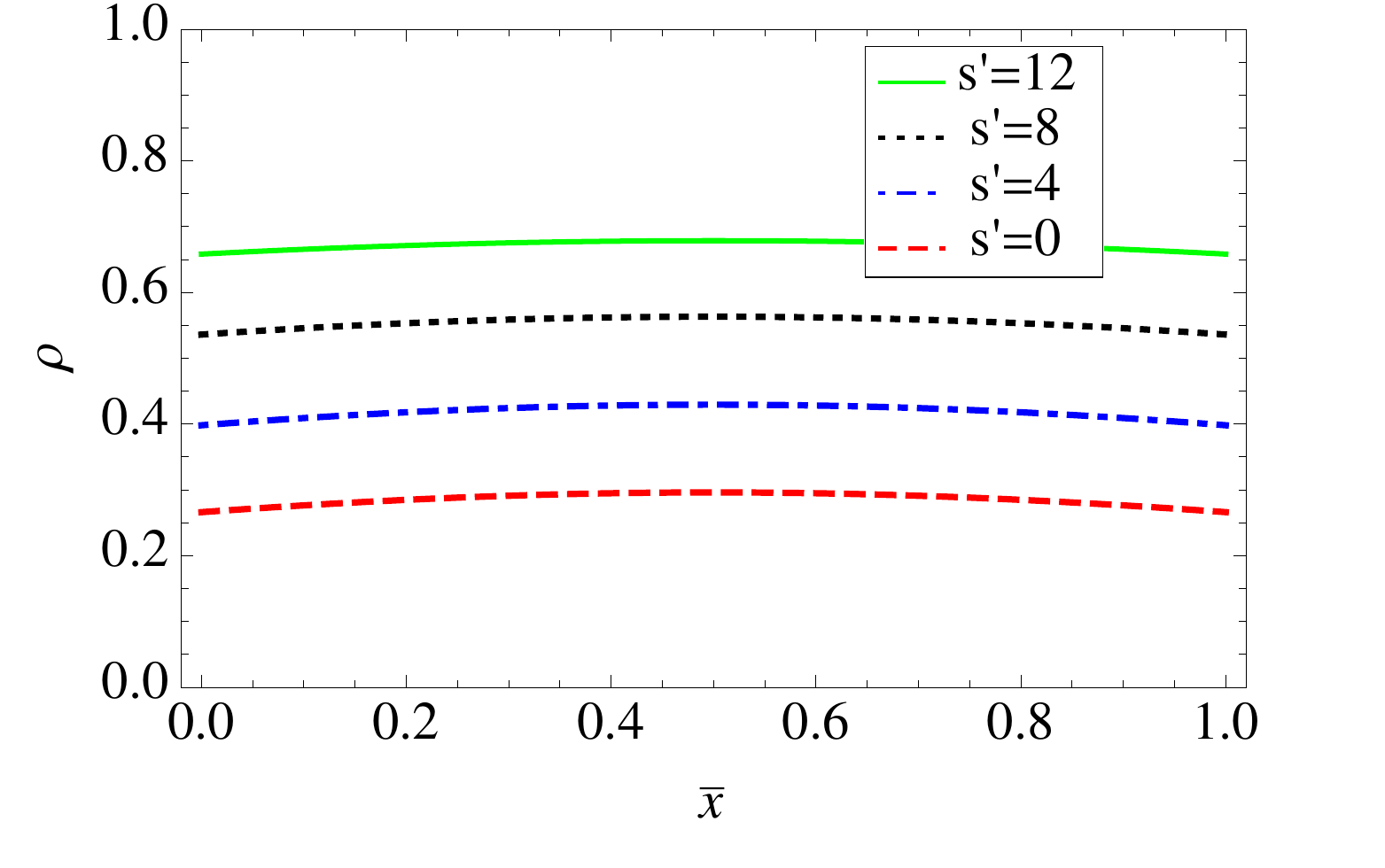}
\caption{(Color online) Estimate of the ratio $\rho=T_{\rm u}/T_{\rm fix1}$ as a function of $\bar x$ for $m'=0.2$, $x_{\rm u}=x_*=1/2$, $\Omega=100$ and $s'=0$, 4, 8, and 12 from bottom to top, calculated as explained in the main text on the basis of the quasi-stationary distribution.}
\label{fig:rho}
\end{figure}
%
%
A numerical study of the estimate of $\rho$ according to Eq.~\reff{rho_eq} is reported in Fig.~\ref{fig:rho} as a function of $\bar x$  for $m'=0.2$, $x_{\rm u}=x_*=1/2$, $\Omega=100$, and for various values of $s'$.
In particular, $\rho$ turns out to increase uniformly as $s'$ increases, which indicates that $T_{\rm u}$ grows faster than $T_{\rm fix 1}$ as a function of this parameter.
It can be noticed that, while generically $\rho$ depends on $\bar x$, this dependence becomes increasingly less important as $m'$ decreases.
The typical $\bar x$-independent estimate of $\rho$ (and therefore of $T_{\rm u}$) can be obtained by considering the mean value
\be\label{rho_medio}
 \rho(s',m')=\int_0^1 {\rmd}\bar x\,A(\bar x)\rho(\bar x,s',m'),
\ee
which depends on the \emph{a-priori} distribution $A(\bar x)$ of the frequency $\bar x$. 
However, as we pointed out above, $\rho(\bar x,s',m')$  is approximately independent of $\bar x$ at least for sufficiently small $m'$ and therefore the specific form of $A(\bar x)$ is inconsequential, so that we can set $A\equiv 1$ in Eq.~\reff{rho_medio}.
Figure~\ref{fig:rho2} shows the dependence of $\rho$ on the migration rate $m'$, as obtained from the numerical integration  of Eq.~\reff{rho_medio}  for various values of the selection coefficient $s'$.
We note again that $\rho$ is an increasing function of the selection coefficient $s'$, and this can be heuristically understood from the fact that balancing selection favors the location of the deme frequency $x_i$ around the optimal frequency $x_*$.
Using Eq.~\reff{rho_medio}, it is straightforward to obtain a numerical estimate for the lifetime of the intermediate state as $T_{\rm u}\simeq   \rho T_{\rm fix1}$.
%
%
\begin{figure}[h!]
\centering
\includegraphics[width=1\columnwidth]{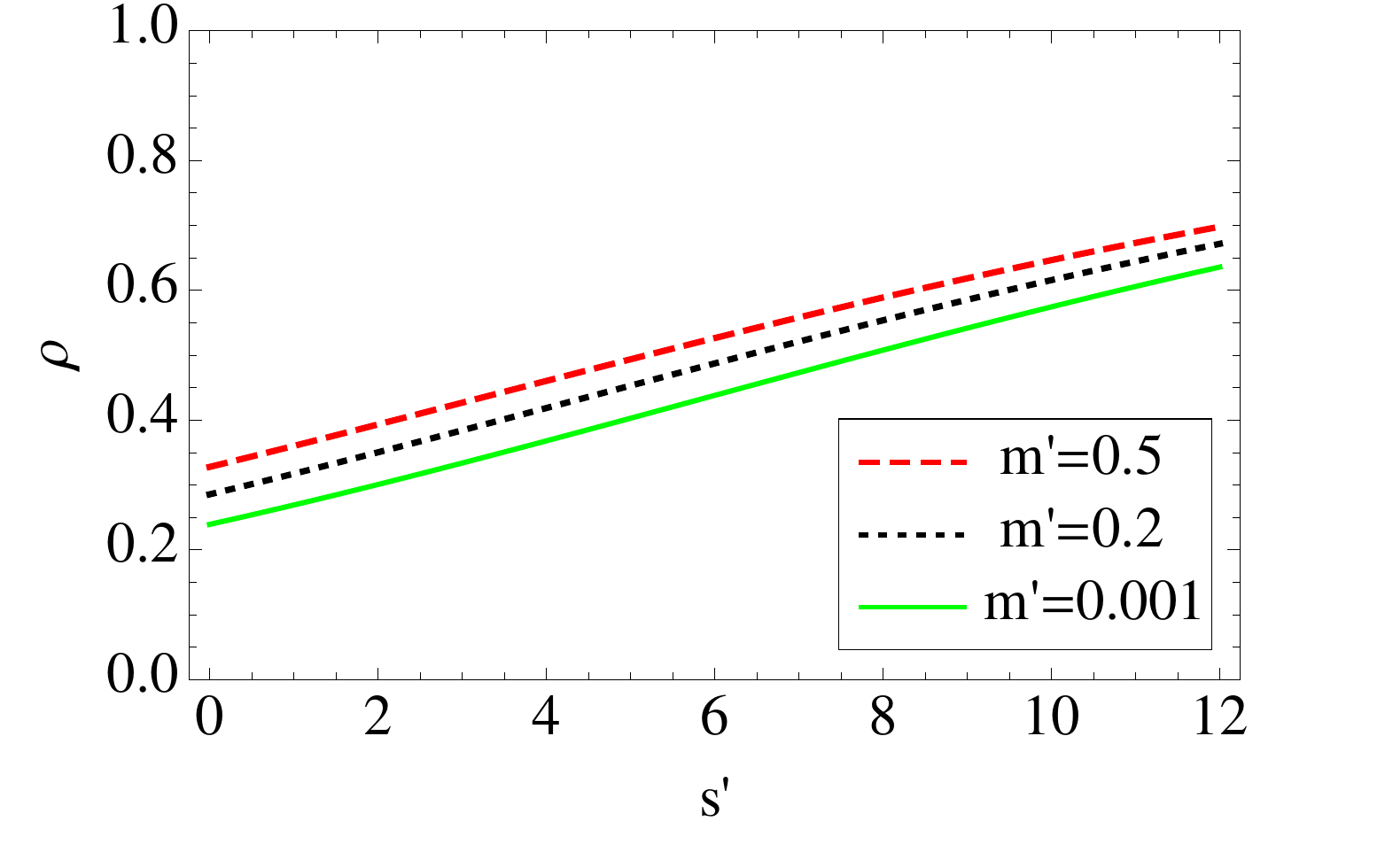}
\caption{(Color online) Mean fraction $ \rho$ of time that a deme spends close to the intermediate state $x_{\rm u}$ as a function of the migration rate $m'$, for  $x_{\rm u}=x_*=1/2$, $\Omega=100$, and $m'=0.001$, 0.2, and 0.5 from bottom to top; these values of $\rho$ have been estimated on the basis of Eq.~\reff{rho_medio} with an uniform \emph{a-priori} distribution $A(\bar x)\equiv 1$.}
\label{fig:rho2}
\end{figure}
%
%

\section{Derivation of the effective equation for $\bar x$ in the voter model with intermediate state}

In Sec.~\ref{sec:voter_int} we introduced a simplified description of the metapopulation consisting of $N$ demes in terms of a voter model with $N$ individuals which can have a definite (0,1) or no definite opinion. Each deme of the original island model is represented by a voter, with opinion 0 or 1 depending on whether the deme has almost fixated at the values $x=0$ or 1 of the frequency $x$ of allele $A$, while individuals with no definite opinion correspond to demes with 
$x = x_{\rm u} \simeq x_*$  
fluctuating in an intermediate long-lived state. The dynamics of the island model can therefore be described at this coarser level by following the evolution of the numbers $N_{\rm u}$, $N_0$, and $N_1 = N - N_{\rm u} - N_0$ of individuals with intermediate opinion, or opinions 0 and 1, respectively.

\subsection{Evolution of $N_{\rm u}$}\label{app-Nu}

Under the assumption that $\bar x = (N_1 + x_{\rm u} N_{\rm u})/N$ is constant (or slowly varying), the behavior of $N_{\rm u}$ is described by a conditional quasi-stationary distribution $P_{\rm qs}(N_{\rm u}|\bar x)$; 
this distribution is the stationary solution of the master equation
\be
 0 = 
 \sum_{n=-1,1,2}\left[W_{N_{\rm u}-n\to N_{\rm u}}P_{N_{\rm u}-n}-W_{N_{\rm u}\to N_{\rm u}+n}P_{N_{\rm u}}\right]
\ee
where, for small $N_{\rm u}/N$ and considering the symmetric case $x_{\rm u}=x_*=1/2$ for simplicity, the rates in Sec.~\ref{sec:voter_int} can be written as
\be
 \begin{split}
  &W_{N_{\rm u}\to N_{\rm u}+1} =m'P[2N(1-P)\bar x(1-\bar x)+N_{\rm u}/2],\\
  &W_{N_{\rm u}\to N_{\rm u}+2} =m'NP^2\bar x(1-\bar x),\\
  &W_{N_{\rm u}\to N_{\rm u}-1} =N_{\rm u}/T_{\rm u},
 \end{split}
\ee
where $P=p(x_{\rm u}|1/\Omega)$ is the probability that a deme with initial frequency $x=1/\Omega$ reaches the intermediate state $x=x_{\rm u}$ before fixation.
Since the rates $W_{N_{\rm u}\to N_{\rm u}'}$ are linear functions of $N_{\rm u}$, the evolution of the mean value  $\langle N_{\rm u}\rangle=\sum_{N_{\rm u}=0}^{\infty}N_{\rm u}P_{\rm qs}(N_{\rm u}|\bar x)$ can be written in closed form: 
\be
 \begin{split}
  & \partial_t\langle N_{\rm u}\rangle=\sum_{ n} nW_{ n}(\langle N_{\rm u}\rangle).
 \end{split}
\ee
The approximate expression reported in Eq.~\reff{N_umedio} can be obtained from the stationary condition $\partial_t\langle N_{\rm u}\rangle=0$, under the assumption of large $N$ and small $m'$.

\subsection{Evolution of $\bar x$}\label{app-xbar}

We study here the evolution of $\bar x$ by considering the fluctuations of $N_{\rm u}$ around its mean given in Eq.~\reff{N_umedio}; 
for the sake of simplicity we focus on the symmetric case $x_*=1/2$, but the discussion below carries over to a generic value of $x_*$.
Because of the presence of demes in the intermediate state, the value of $\bar x$ receives a contribution of the form
\be
 \bar x=\frac{N_1}{N}+\frac{N_{\rm u}}{2N},
\ee
and therefore
\be\label{N0N1_x}
 \begin{split}
  N_1 &=N\bar x-N_{\rm u}/2,\\
  N_0 &=N(1-\bar x)-N_{\rm u}/2.
 \end{split}
\ee
These relations can now be used in order to express the rates $W_{A,\ldots,D}$ in Eq.~\reff{rates} as functions of $N_{\rm u}$. For large $\Omega$ 
these expressions can be further simplified by taking into account that $P$, $Q\propto 1/\Omega$ for $s'$ not too large (we recall here that $P=p(x_{\rm u}|\Omega^{-1})$, $Q=p(x_{\rm u}|1-\Omega^{-1})$ --- see Sec.\ref{sec:voter_int} --- with $p$ given in Eq.~\reff{pfix}),
and that $T_{\rm u}\propto T_{\rm fix 1}\propto\Omega$; one eventually finds
\be\label{rates2}
 \begin{split}
  W_A &=m'NP\left\{ \bar x(1-\bar x)-\frac{N_{\rm u}}{N}\frac{\bar x}{2} \right\}+O(1/\Omega^2),\\
  W_B &=m'NP\left\{ \bar x(1-\bar x)-\frac{N_{\rm u}}{N}\frac{1-\bar x}{2} \right\}+O(1/\Omega^2),\\
  W_C &=O(1/\Omega^2),\\[2mm]
  W_D &=W_E=N_{\rm u}/(2T_{\rm u}).
 \end{split}
\ee
As discussed in Sec.~\ref{sec:effective_eq}, the evolution of $\bar x$ can be described by an effective Langevin equation (diffusion approximation, see Appendix \ref{app-diffusion}) in the large $N$ limit,
\be\label{Langevin}
 \dot {\bar x}=M_{\rm vi}(\bar x)+\sqrt{V_{\rm vi}(\bar x)}\,\eta(t),
\ee
where the drift and variance are given by 
\be
 \begin{split}
  M_{\rm vi}(\bar x) &=\sum_{i=A,B,C,D,E}W_i\Delta\bar x_i,\\
  V_{\rm vi}(\bar x) &=\sum_{i=A,B,C,D,E}W_i(\Delta\bar x_i)^2,\\
 \end{split}
\ee
and the relevant increments are $\Delta \bar x_A=\Delta\bar x_E=1/(2N)$ and $\Delta \bar x_B=\Delta\bar x_D=-1/(2N)$.
Then, by replacing $N_{\rm u}$ with its mean $\langle N_{\rm u} \rangle$ reported in Eq.~\reff{N_umedio}, we eventually obtain the result anticipated in Eq.~\reff{Langevin_vi}.


\begin{thebibliography}{99}
\bibitem{notaMoran}
In order to recover for large $\Omega$ the same behavior as the previously described Wright-Fisher model, we have to choose a duration $\delta t=2\tau_g/\Omega$ for the Moran time step and to replace the rates $m$ and $s$ with $2m$ and $2s$ respectively (see Ref.~\cite{prl}).
\bibitem{notatime scale}
More precisely, the time scale separation emerges when $N\gg1$ and either $N s'<1+1/m'$, or 
$$
 \left\{
 \begin{array}{l}
   N s'>1+1/m'\\
   s_{\rm e}/m\ll 1
 \end{array}
 \right. ,
$$
where $s_{\rm e}=s/\left[\left(1+\frac{1}{m'}\right)\left(1+\frac{1}{2m'}\right)\right]$ is an effective selection coefficient.
\bibitem{notavoter2} 
This is the rate associated with the probability $P^{\geq 1}(\delta t)$ that at least one individual is exchanged between one of the possible $N(N-1)/2$ pairs of demes of the population in the small time interval $\delta t$, thus establishing an ``interaction'' between the corresponding voters. In this time interval, each pair of demes exchanges a mean number $\varepsilon \equiv m\Omega\delta t/(N-1)$ of individuals and, according to the Poisson counting, the probability that there is no exchange is $\rme^{-\varepsilon} \simeq 1-\varepsilon$ in that pair and therefore $\simeq (1-\varepsilon)^{N(N-1)/2} \simeq 1 - m'N\delta t/2$ in all possible pairs. This probability equals $1-P^{\geq 1}(\delta t)$ and thus $P^{\geq 1}(\delta t) \simeq m'N\delta t/2$, which results in the rate reported in Eq.~\reff{rate vot}.
\bibitem{nota-stato-iniziale}
Since the number of uncertain voters is supposed to be small ($N_{\rm u}\ll N$), the state $\bar x=1/2$ of the voter model can be essentially realized only by having $(N_0=N/2,N_1=N/2)$  as the initial condition, which is, in fact, the same initial condition as in Ref.~\cite{prl} and for $x_*\sim 1/2$ is also a good approximation of the metastable state into which the system quickly relaxes.
\end{thebibliography}
\end{document}